\documentclass{emulateapj}

\slugcomment{}
\usepackage{graphicx}
\usepackage{longtable}
\usepackage{amsmath}

\shorttitle{High-Redshift Galaxy Formation at the Highest Luminosities}
\shortauthors{Lee et al. }
\begin{document}
\def\hh{\, h^{-1}}
\def\c1334{C {\sc ii} $\lambda$1334}
\def\si1393{Si {\sc iv} $\lambda$1393}
\def\si1260{Si {\sc ii} $\lambda$1260}
\def\osi1303{O {\sc i}+Si {\sc ii} $\lambda$1303}
\title{Probing  High-Redshift Galaxy Formation at the Highest Luminosities: New Insights from DEIMOS Spectroscopy}
\author{Kyoung-Soo Lee\altaffilmark{1}, Arjun Dey\altaffilmark{2},  Michael C.~Cooper\altaffilmark{3}, Naveen Reddy\altaffilmark{4}, Buell T. Jannuzi\altaffilmark{5}}
\altaffiltext{1}{Department of Physics, Purdue University, 525 Northwestern Avenue, West Lafayette, IN 47907}
\altaffiltext{2}{National Optical Astronomy Observatory, Tucson, AZ 85726}
\altaffiltext{3}{Department of Physics and Astronomy, University of California, Irvine, CA 92697}
\altaffiltext{4}{Department of Physics and Astronomy, University of California, Riverside, 900 University Avenue, Riverside, CA 92521}
\altaffiltext{5}{Steward Observatory, University of Arizona, Tucson, AZ 85721}

\begin{abstract}
We present Keck DEIMOS spectroscopic observations of the most UV-luminous star-forming galaxies at redshifts $3.2<z<4.6$. Our sample, selected in the Bo\"otes field of the NOAO Deep Wide-Field Survey, contains galaxies with luminosities of $L^* \lesssim L_{\rm{UV}} \lesssim 7L^*$ and is one of the largest samples to date of the most UV-luminous galaxies at these redshifts.  Our spectroscopic data confirm 41 candidates as star-forming galaxies at $3.2<z<4.6$ and validate the relatively clean selection of the photometric candidates with a contamination rate of 11\%-28\%. We find that the fraction of Ly$\alpha$ emitting galaxies increases with decreasing UV luminosity. None of the 12 galaxies with $M_{\rm{UV}}<-22$ (i.e., $L_{\rm{UV}}>3L^*$) exhibit strong Ly$\alpha$ emission. We find strong evidence of large-scale outflows, transporting the neutral/ionized gas in the interstellar medium away from the galaxy. Galaxies exhibiting both interstellar absorption and Ly$\alpha$ emission lines show a significant offset between the two features, with the relative velocity of $200-1150$~km~s$^{-1}$. We find tentative evidence that this measure of the outflow velocity increases with UV luminosity and/or stellar mass. The luminosity- and mass-dependent outflow strengths suggest that the efficiency of feedback and  enrichment of the surrounding medium depend on these galaxy parameters. We also stack the individual spectra to construct composite spectra of the absorption-line-only and Ly$\alpha$-emitting subsets of the UV luminous galaxies at $z\simeq  3.7$. The composite spectra are very similar to those of lower-redshift and lower-luminosity LLyman break galaxy (LBG) samples, but with some subtle differences.  Analyses of the composite spectra suggest that the UV luminous LBGs at $z\simeq 3.7$ may have a higher covering fraction of absorbing gas, and may be older (or have had more prolonged star formation histories) than their lower-redshift and lower-luminosity counterparts. In addition, we have discovered that five galaxies in the sample belong to a massive overdensity at $z=3.78$. Finally, two galaxies each show two distinct sets of interstellar absorption features. The latter may be a sign of a final stage of major merger, or clumpy disk formation. These systems are not expected in our sample: their presence implies that frequency of such sources among our luminous $z\simeq 3.7$ LBGs may be an order of magnitude higher than in lower redshift and lower luminosity samples. 

\end{abstract}
  
\keywords{cosmology:observations -- galaxies:distances and redshifts -- galaxies:evolution -- galaxies:formation}

\section{Introduction}

Deep multi-wavelength space- and ground-based imaging surveys have enabled the identification and study of large numbers of high-redshift galaxies \citep[e.g.,][]{steideletal03,steideletal04,mauro04a}. These studies have made significant strides in characterizing various global statistics of the high-redshift galaxy population, including their UV luminosity function, stellar mass function, and clustering properties \citep[e.g.,][]{steidel99,mauro04b,bouwens07,reddy09,ouchi04a,ouchi04b,lee06,lee09,lee12a,gonzalez11}. In addition, the overall distribution of dust content, stellar population ages, and sizes have been determined at different cosmic epochs \citep{ferguson04, bouwens04c, bouwens09, bouwens12, stark09, reddy06a, reddy12_herschel, lee12b,finkelstein12}. 

Nevertheless, several questions remain about the origin and evolution of these galaxies. For example, what physical processes govern the star-formation histories of these galaxies? Do the star-formation, assembly and chemical evolutionary histories vary as function of galaxy luminosity, halo and stellar mass, and environment? If they do, what drives these differences? 

While useful constraints have been placed on some of these questions based on statistical studies \citep[e.g.,][]{steidel05,lee11,lee12a,papovich11,reddy12_sfh}, the deep spectroscopy necessary to measure the age and metallicity of the stellar populations, interstellar abundances, and large-scale outflows, has been lacking \citep[but see][for review]{giavalisco02,shapley11}. The main challenge has been the faintness of most high-redshift galaxies, which precludes high signal-to-noise ratio (S/N) spectroscopic studies. The usefulness of deep spectroscopy in shedding light on the chemical, kinematic, and geometrical details within the galaxy has been amply demonstrated in a few studies of very rare, bright galaxies \citep[either intrinsically luminous or gravitationally amplified; e.g., MS1512-cB58:][]{pettini00,pettini02, steidel10}, and also by studies of high S/N composite spectra created by averaging many low-S/N spectra \citep[e.g.,][]{shapley03, steidel01, steidel10,jones12}. These pioneering studies have shown that galactic outflows are powerful enough to drive out as much gas as that consumed by the star-formation at velocities close to the escape velocity, thereby effectively enriching the circum-galactic medium. \citet{shapley03} showed that the interstellar medium (ISM, hereafter) in these galaxies must be clumpy based on the observed anti-correlation between the Ly$\alpha$ and interstellar line equivalent widths. When a galaxy is observed through optically thinner regions, more Ly$\alpha$ photons escape while less interstellar absorption will occur. 

Most of these results pertain to galaxies in a relatively narrow range of UV luminosity ($\approx L^*$) and redshift (typically $z\simeq 2-3$). This luminosity and redshift ranges represent a ``sweet spot'' for efficient studies of relatively large numbers of UV-luminous galaxies, since the space density of high-redshift galaxies falls precipitously at $L\gtrsim L^*$ while the success rate of spectroscopic identification declines to lower luminosities and higher redshifts.  However, several studies have begun to suggest that galaxies of different luminosities may exhibit different characteristics \citep[e.g., the fraction of Ly$\alpha$ emitting galaxies or  the stellar-mass-to-UV-light ratio: see][]{stark10,lee12a}. It is therefore important to investigate the full luminosity range of high-redshift galaxies and determine how {\it representative} the current physical picture, offered by spectroscopic efforts of $L\gtrsim L^*$ galaxies, is to more- or less-luminous galaxies and/or at higher redshift.
Widening the dynamic range will also improve our understanding of the physical processes of the feedback and mass assembly histories within different dark matter halo environments \citep[e.g.,][]{ouchi04b,lee06,hildebrandt07}. 

In this paper, we present a pilot study of spectroscopic observations of higher-redshift ($3.2<z<4.6$), higher-luminosity ($L^* \lesssim L \lesssim 7L^*$) galaxies with the primary aim of  critically assessing our current understanding of galaxy formation at high redshift. %

We use the {\it WMAP7} cosmology $(\Omega, \Omega_\Lambda, \sigma_8, h_{100})=(0.27, 0.73, 0.8, 0.7)$ \citep{wmap7}. Adopting the {\it WMAP9} or {\it Planck} cosmology \citep[][respectively]{wmap9, planck13} would change  angular sizes or luminosities by less than 1\%. Magnitudes are given in the AB system \citep{oke83} unless noted otherwise.

\section{Data and Photometric Selection of Galaxies at $z\simeq 3.7$}\label{data}
Our photometric candidates are selected from the Bo\"otes field of the NOAO Deep Wide-Field Survey \citep[NDWFS;][]{jannuzid99}. Briefly, the NDWFS $B_WRI$ band data were obtained using the Mosaic camera on the Mayall 4m Telescope, and consist of 27 separate pointings covering a contiguous 9.3 deg$^2$ area. 

For the spectroscopic target selection, we focused on three pointings obtained under the best observing conditions. We chose pointings where: (1) all three photometric bands have seeing better than 1\arcsec\ and within 0.1\arcsec\ of one another; and (2) the 50\% completeness limit in the three bands are deeper than $B_W > 26.8$, $R>25.8$, and $I>25.1$  (Vega)\footnote{The NDWFS images are calibrated in Vega magnitudes. The correction to be applied to convert to the AB system is 0.019, 0.215, and 0.459 mag for the $B_W$, $R$ and $I$-band, respectively; e.g., $I_{\rm{AB}}=I_{\rm{Vega}}+0.459$.}. The completeness limit was determined by adding artificial stellar objects (convolved with a Moffat profile matched to the measured seeing) to the images and recovering them with SExtractor using the identical parameters as the real data (tabulated on the survey web page).\footnote{{\tt http://www.noao.edu/noao/noaodeep/DR3/dr3-descr.html}}  These requirements allow robust photometric color selection without degrading the images by convolving them to a common seeing, and a maximum surface density within the multi-object masks down to $I\approx 25.0$. 

\begin{figure}
\epsscale{1.0}
\plotone{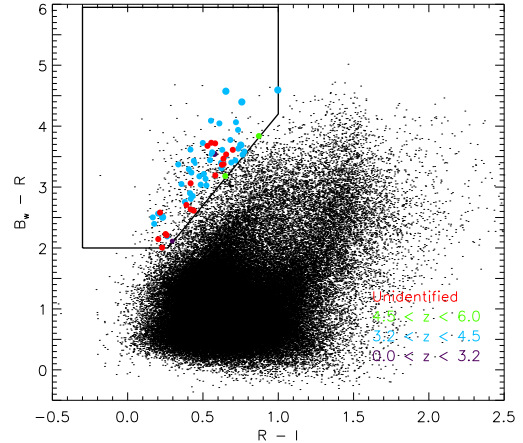}
\caption[ccd_spec]
{
Locations of the NDWFS sources on the $B_W-R$-$R-I$ plane are shown together with the Lyman-break color selection criteria outlined by the black solid line. The colors are based on Vega magnitudes. The spectroscopic targets are overlaid in color showing the interlopers ($z<3.2$; dark purple), LBGs at $3.2<z<4.2$ (blue), LBGs at $z>4.2$ (green), and the sources that could not be identified (red). 
}
\label{ccd_spec}
\end{figure}

Our photometric sample is selected by applying a Lyman-break color-selection technique to the $B_WRI$ data  \citep[e.g.,][]{steidel99, mauro04b, bouwens07}. The Lyman-break technique is designed to identify a UV bright star-forming galaxy  (or Lyman break galaxy, LBG, hereafter)  with a strong spectral break (at a rest-frame wavelength $\lambda\le 1216$ \AA) resulting from absorption by the intervening Ly$\alpha$ forest. At $3.3<z<4.3$, this break lies between the $B_W$ and $R$ bands. The color criteria (in Vega magnitudes) used to select the candidates are identical to that used for a larger sample of photometric candidates described in \citet{lee11}. 
\begin{eqnarray}
(B_W-R) \ge 1.2+3.0 \times (R-I) ~~~~~ (B_W-R) \ge 2.0~~ \nonumber\\
 (R-I) \ge -0.3 ~~~~~ \rm{S/N}(R) \ge 3 ~~~~~ \rm{S/N}(I) \ge 7~~~~~~~
\end{eqnarray}
The total number of sources satisfying the color selection criteria are 738. In Figure \ref{ccd_spec}, we show the color locations of all sources together with the selection window. The spectroscopic targets are shown as colored dots; these are discussed further in \S\ref{keck}. 

\begin{figure*}
\epsscale{1.}
\plottwo{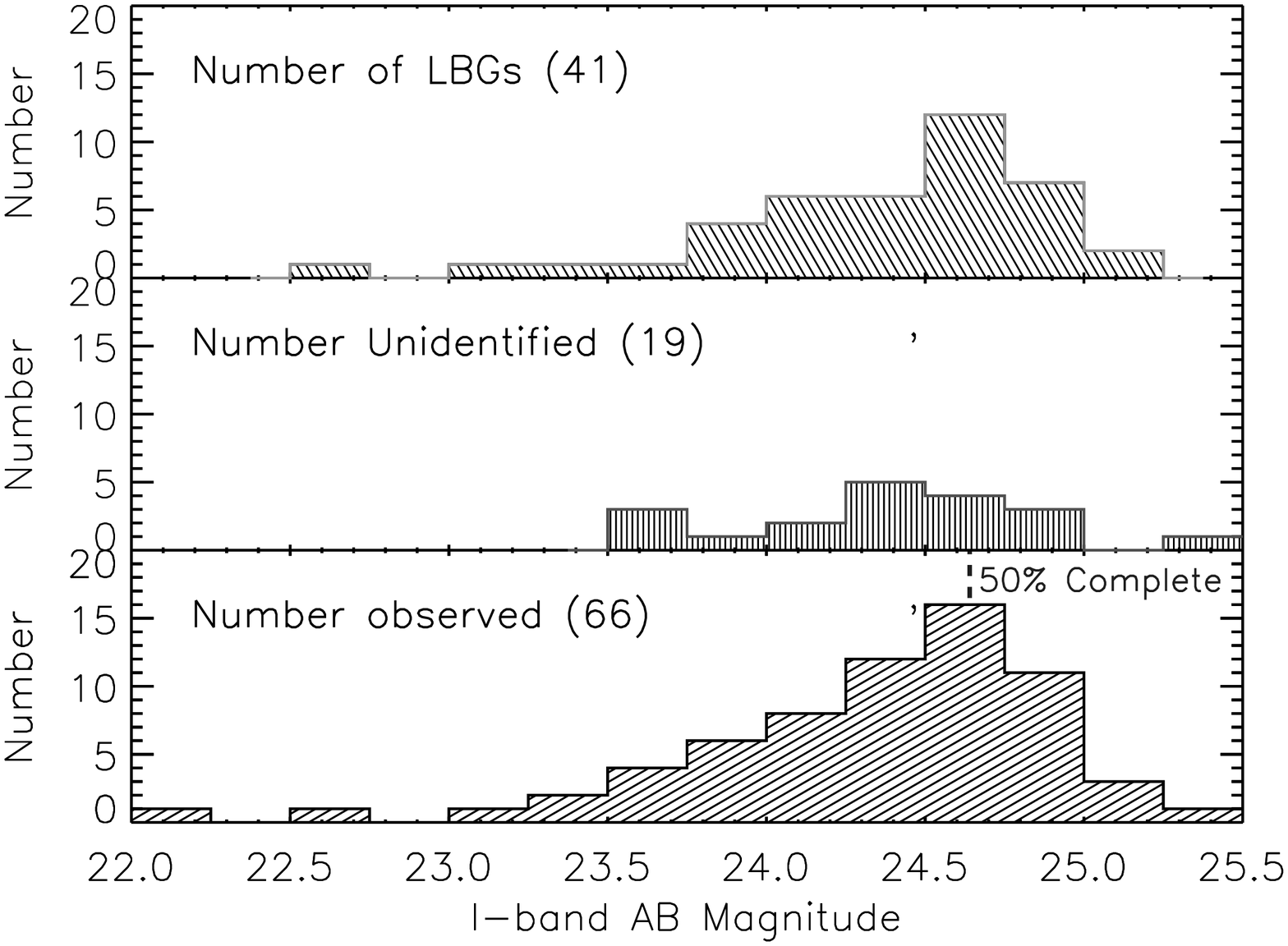}{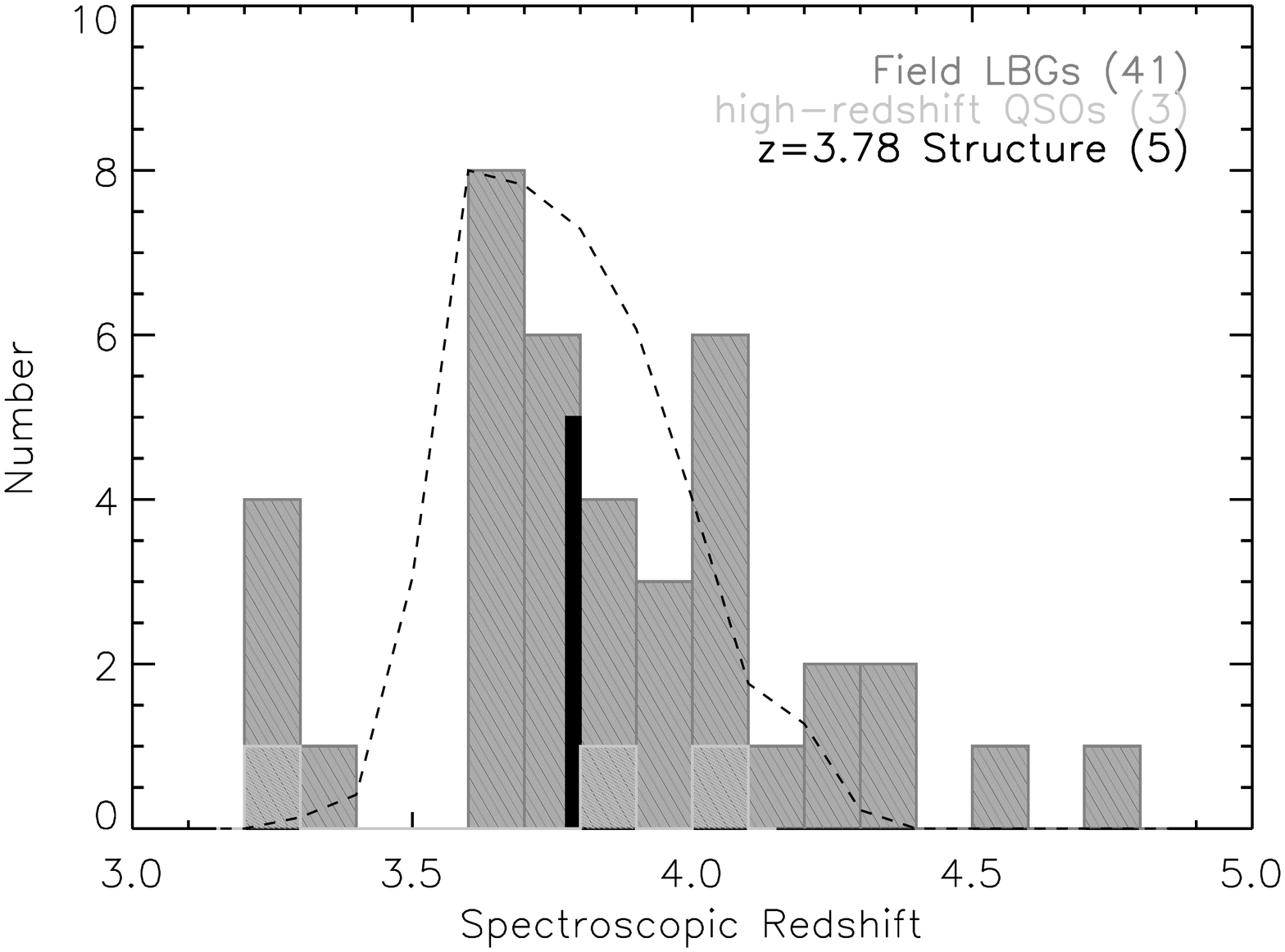}
\caption[spec]
{
Left: the magnitude distribution of our targets (66), confirmed LBGs (41) at $z=3.2-4.7$, and unidentified sources (20) are shown as a function of $I$-band magnitude. The magnitude at which the source detection is 50\% complete is indicated as dashed line on top. 
 Right: the redshift distribution of the spectroscopically confirmed LBGs and QSOs in the NDWFS Bo\"otes field. Also shown are the redshift distribution of three QSOs and five galaxies in the overdensity at $z=3.78$. The expected redshift distribution from our photometric simulations is shown in dotted line. 
}
\label{spec}
\end{figure*}

\section{Keck DEIMOS Observations}\label{keck}

Spectroscopic observations of our LBG candidates were carried out on the night of May 10 2010 using the DEep Imaging Multi-Object Spectrograph (DEIMOS) on the 10-m Keck II telescope \citep{deimos_ref}. We used 6000 line mm$^{-1}$ grating blazed at 7500\AA, yielding a spectral coverage of 4300-9600\AA\ at the dispersion of 0.65\AA~ pixel$^{-1}$. 

We designed three masks (BDm\_01, BDm\_02, and BDm\_03) covering a total of 66 candidates. The mask locations were selected with the aim of maximizing the number of LBG candidates within each DEIMOS field of view. The photometric candidates were put in two categories as ``primary'' ($\rm{S/N}(I)\geq 10$) and ``secondary'' ($7\leq \rm{S/N}(I)< 10$) targets, with the primary targets driving the DEIMOS field placement. Table \ref{tbl_1} summarizes the basic properties of each mask. 
The slit width was 1\arcsec.2 in all three masks. The effective spectral resolution estimated from the sky lines is 3.2\AA. Each mask was observed with $2.0-2.5$ hr in good conditions, with seeing between 0\arcsec.6 and 0\arcsec.8 throughout the night.  

\begin{deluxetable*}{lcccccc}
\tabletypesize{\scriptsize}
\tablewidth{0pc}
\tablecaption{Summary of DEIMOS Multi-object Slit Masks}
\tablehead{
\colhead{   Mask} &
\colhead{Field\tablenotemark{a}} &
\colhead{R.A.\tablenotemark{b}} &
\colhead{Decl.\tablenotemark{b}} &
\colhead{$N_{\rm{primary}}$} &
\colhead{$N_{\rm{secondary}}$} &
\colhead{$t_{\rm{exp}}$ (hr)}}
\startdata
BDm\_01 & NDWFSJ1426p3236 & 14:27:10.30 & 32:34:11.0 & 20 & 9 & 2.50\\
BDm\_02 & NDWFSJ1431p3236 &14:31:26.41  & 32:25:40.8 & 14 & 0 & 2.25\\
BDm\_03 & NDWFSJ1437p3347 &14:37:34.37  & 33:36:36.1 & 16 & 7 & 1.99\\
All & --- &---& ---& 50 & 16
\enddata
\label{tbl_1}
\tablenotetext{a}{The name of the fields in the NDWFS survey Web site}
\tablenotetext{b}{The J2000 coordinates of the center of each pointing}
\end{deluxetable*}

The data were reduced using the spec2d IDL pipeline developed for the DEIMOS instrument \citep{spec2d_ref,cooper12}. The wavelength calibration was carried out using observations of Ne+Ar+Kr+Xe arc lamps. The two-dimensional sky-subtracted spectra were inspected visually to validate the automatic extraction of the one-dimensional spectra by the DEIMOS pipeline. There are a few cases where  another galaxy fell into the slit and affected the sky subtraction, or the galaxy was located too close to the edge of the slit. In these cases, re-extraction using the IRAF routine {\tt noao.twodspec.apall} significantly improved the quality of the one-dimensional spectra.

\begin{deluxetable*}{lcccccccccc}
\tabletypesize{\scriptsize}
\tablewidth{0pc}
\tablecaption{The Summary of the Spectroscopic Sample}
\tablehead{
\colhead{ID} &
\colhead{R.A.} &
\colhead{Decl.} &
\colhead{$I$ mag} &
\colhead{Class\tablenotemark{a}} &
\colhead{$z_{{\rm Ly\alpha}}$\tablenotemark{b} } &
\colhead{$z_{{\rm IS}}$\tablenotemark{c} } &
\colhead{$W_{0,\rm{Ly}\alpha}$\tablenotemark{d}} &
\colhead{UV~continuum\tablenotemark{e}} &
\colhead{FLAG([3.6~$\mu m$])\tablenotemark{f}} }
\startdata
BD13355 & 14:27:15.864 & 32:32:34.66 & 24.27 &P& -- & 0.373 & -- & Y& N\\
BD6909 & 14:27:12.180 & 32:25:58.69 & 25.08 &P& -- & 0.631 & -- &Y& N\\
BD28913 & 14:27:08.131 & 32:41:08.92 & 23.38 &P& -- & 3.289 & [QSO] &Y&Y\\
BD106241 & 14:31:06.041 & 32:30:11.12 & 22.24 &P& -- & 3.835 & [QSO] &Y& Y\\
BD10334 & 14:27:04.982 & 32:29:30.84 & 23.78 &P& 4.041 & -- & [QSO] &Y&Y\\
BD97772\tablenotemark{g} & 14:31:47.057 & 32:22:56.21 & 25.06 & P&3.782  & 3.773 & 5.1&Y& N\\
BD98176\tablenotemark{g} & 14:31:48.811 & 32:23:23.10 & 24.61 & P&3.784  & 3.775 & 14.2 &Y& N\\
BD99073 & 14:31:37.656 & 32:24:17.39 & 24.16 & P& 3.730  & 3.717 & 0.9 &Y& ?\\
BD99847\tablenotemark{g} & 14:31:37.711 & 32:25:03.07 & 24.54 & P& 3.783   & --& 21.3 &Y&Y\\
BD100006\tablenotemark{g} & 14:31:54.209 & 32:25:11.06 & 23.76 & P& 3.787 &  3.779 & 10.9 &Y& Y\\
BD100902\tablenotemark{g} & 14:31:37.001 & 32:25:57.54 & 24.45 & P&  3.782  & 3.777 & 24.9 &Y& N\\
BD102774 & 14:31:45.130 & 32:27:28.91 & 24.10 &P& -- & 3.753 & -3.8 &Y&Y\\
BD98449 & 14:31:42.230 &32:23:40.67 & 24.75  & P& 4.312 & -- & 26.3 &Y& N\\
BD100141 & 14:31:40.255 & 32:25:18.44 & 24.72 &P& -- & 3.992 & -8.0 &Y&Y\\
BD7645 & 14:27:14.009 & 32:26:45.17 & 24.05 &P& -- & 3.685,3.679\tablenotemark{h}  &-7.2 &Y& Y\\
BD8040 & 14:27:11.386 & 32:27:08.03 & 25.11 &S& 3.671  &  3.666 & 15.1&Y& N\\
BD9126 & 14:27:15.394 & 32:28:15.20 & 24.59 &S& 4.091  & -- & $>19.8$ & N & N\\
BD10059 &14:27:04.711 & 32:29:13.56 & 24.85 &S& -- & 3.240 & -8.8 &Y& Y\\
BD10252 & 14:27:01.975 & 32:29:25.30 & 23.80 &P& 4.073 &  4.062 & 9.6 &Y&Y\\
BD11113 & 14:27:08.621 & 32:30:21.67 &24.26 &P & 3.659 &  3.653 & 5.9 &Y& N\\
BD11352 & 14:27:08.940 & 32:30:35.24 & 24.20 &P& -- & 3.742 & -5.2 &Y&Y\\
  BD15308 & 14:27:04.745 & 32:34:27.66 & 23.90 &P & -- & 3.301 & -7.6 &Y&Y\\
BD13266 & 14:27:09.893 & 32:32:28.79 & 23.33 &P& -- & 4.062 & -4.2 &Y&Y\\
BD18449 & 14:26:57.403 & 32:38:19.03 &22.56 & P& 3.744 &  3.726,3.744\tablenotemark{h} & -2.6 &Y&Y\\
BD19284 & 14:27:00.732 & 32:39:24.66 & 24.31 &S& -- & 3.260 & -3.7 &Y&Y \\
BD28630 & 14:27:05.933 & 32:41:23.46 & 24.98 &S & 3.844  & 3.839 & -- &Y& N\\
BD29677 & 14:26:59.578 & 32:40:31.76 & 24.76 &S& 4.205  & 4.195& 33.2 &Y&Y\\
BD8859 & 14:27:05.875 & 32:27:56.74& 23.25 &P & -- & 4.099 & -3.8 &Y&Y\\
BD12027 &14:26:56.448 & 32:31:14.81 & 24.74 &P& -- & 4.070 & -6.1&Y& Y\\
BD12510 & 14:27:04.061 & 32:31:44.15 &24.16 & P& 4.717  &  4.704  & 53.9 &Y&Y\\
BD13763 & 14:27:13.939 & 32:32:56.15 &24.80 & P&3.672 & 3.665 &6.7 &Y&?\\
BD14778 &14:27:06.732 &32:33:59.69&24.70& P&3.987 &  3.980 &-0.5 &Y& ?\\
BD235717 &14:37:02.976 & 33:32:30.37 & 24.58& S& 3.826   & 3.814&31.9 &Y& N\\
BD236105 & 14:37:18.703 & 33:32:50.39 & 24.79 & S&4.219 & -- &$>18.3$ &N& N\\
BD239105 & 14:37:24.862 & 33:35:34.62 & 23.96 &P & -- & 3.900 & -13.4 &Y&Y\\
BD239303 & 14:37:38.177 & 33:35:43.12 & 24.86 &S & -- & 3.828 &-1.9 &Y&Y\\
BD239473 & 14:37:52.464 & 33:35:50.75&24.59 &P & -- & 3.690 & -2.5 &Y& ?\\
BD239555 & 14:37:54.758 &  33:35:55.39 & 24.75 &P& -- & 3.230 &-1.2 &Y&Y\\
BD240454 & 14:37:48.238 & 33:36:46.37 &24.49 &P& -- & 3.663 & -1.1 &Y&Y\\
BD240992 & 14:37:44.093 & 33:37:19.88 & 24.61&P & -- & 3.650 & -5.9 &Y& N\\
BD244504 & 14:37:58.382  & 33:41:06.86 & 24.29 &P& -- & 3.724 & -1.9 &Y& N\\
BD235538 & 14:37:11.563 & 33:32:18.46 & 24.22 &S& 4.327  & 4.319 & 48.7 &Y&Y\\
BD239525 & 14:37:57.312 & 33:35:52.80 &23.61 &P& -- & 3.750 & -0.7 &Y& N\\
BD242042 & 14:37:36.074  & 33:38:28.61 & 24.41 &S& -- & 4.506 & -4.5 &Y&Y\\
BD243232 & 14:37:57.113 &  33:39:43.63 & 24.57 &P & -- & 4.201 & -4.2 &Y&Y\\
BD244796 & 14:37:58.128 & 33:41:23.46 & 24.56 &P&  3.621 &  3.610 & 14.64 &Y& N\\
\enddata
\label{tbl_2}
\tablenotetext{a}{``P'' is for primary and ''S'' is for secondary targets, based on the S/N in the detection band}
\tablenotetext{b}{The Ly$\alpha$ redshift is determined from the peak of the Ly$\alpha$ emission}
\tablenotetext{c}{The interstellar redshift is determined from the peak of the interstellar absorption line}
\tablenotetext{d}{The rest-frame equivalent width of Ly$\alpha$, only measured for LBGs}
\tablenotetext{e}{The UV continuum detected in the spectrum?}
\tablenotetext{f}{{\em Spitzer} IRAC [3.6~$\mu m$] band detection: ``Y'' for detected, ``N'' for undetected, and ``?'' for uncertain}
\tablenotetext{g}{The galaxies in the $z=3.78$ structure (see \S\ref{protocluster})}
\tablenotetext{h}{The galaxies with multiple interstellar redshifts (see \S\ref{double_abs})}
\end{deluxetable*}

Redshifts were determined by visual examination from both one- and two-dimensional spectra. We used the SpecPro software \citep{specpro_ref} to check the spectra simultaneously with the available imaging data. Redshift identification is made by looking for the presence of Ly$\alpha$ emission,  interstellar absorption lines, and the continuum break due to the Lyman alpha forest.
Of the 66 $B_W$-band dropout candidates observed during the run, we have determined redshifts for 46 sources. In Table \ref{tbl_2}, we list the basic properties of the 46 galaxies/QSOs including their coordinates, redshifts, and magnitudes. 

Of the 46 spectroscopic identifications, 41 sources are galaxies at the expected redshift range, 2 are low-redshift interlopers at $z=0.373$ and $0.631$ (identified via [O\,{\sc ii}] 3727  and H$\alpha$ emission lines), and 3 are QSOs at $z= 3.289, 3.835, 4.041$.  Of the remaining 20 sources for which we were unable to determine redshifts, 2 show  featureless red continua, indicating that they are likely dusty galaxies at $z<2$. An additional 5 galaxies show a clear continuum break consistent with being an LBG, but the S/N is too poor to identify emission or absorption features to confirm (or measure precise) redshifts. Finally, no information could be obtained on 2 sources because the slitlet was contaminated and therefore masked out during the reduction.  Hence, the estimated contamination rate from the spectroscopic identifications is 11\% (5/46), where we have included the $3<z<4$ QSOs as contaminants. The lower limit on the contamination rate for the full sample is also 11\% ($=(5+2)/(66-2)$) where the total of 64 galaxies excludes the two sources which were masked out. In the most pessimistic case where all of the unidentified sources are not LBGs, the contamination rate is 28\%; $1-(41+5)/(66-2)$. The magnitude distributions of our targets, confirmed LBGs, and unidentified sources are presented in Figure \ref{spec} (left). 

For the majority of confirmed LBGs (36/41; i.e., 88\%), a clear UV continuum and at least two interstellar lines were detected with sufficient S/N to determine interstellar redshifts (Table~\ref{tbl_2}). The remaining 5 LBGs are identified only by their Ly$\alpha$ emission. Of those 5,  we detect  a UV continuum in three galaxies but at a low S/N that we could not measure interstellar redshifts.  Therefore, the confirmed LBGs at $z\leq 4.2$ represent approximately a (continuum-)flux-limited sample minimally affected by the spectroscopic bias which preferentially identifies line emitters. 

Figure \ref{spec} (right) shows the redshift distribution of all spectroscopically identified sources at $z>3$, including 41 LBGs and 3 high-redshift QSOs. The observed high-redshift tail (at $z>4.2$) can be explained by the spectroscopic selection effect that favors redshift identification for Ly$\alpha$ emitting sources. Of the seven sources at $z>4.2$, four are secondary targets, detected at a low S/N in the $I$-band, and five are observed with a high Ly$\alpha$ rest-frame EW ($W_0\geq18$\AA). The average EW for those five sources is $W_0=24.4$\AA, much higher than that measured for the full sample ($W_0=-1.2$\AA). Among the measured redshifts are five galaxies lying within a small redshift range ($z=3.782-3.787$) and within 1.5\arcmin\ from each other (all in the mask BDm\_02). The close physical proximity of these five galaxies strongly suggests that they belong to a massive structure (see \S\ref{protocluster}). In consideration of these factors,  the observed $N(z)$ is in a reasonable agreement with the redshift distribution expected from photometric simulations as shown in Figure \ref{spec} (dotted line in the right panel; see Lee et al. 2011 for the details of the simulations), i.e., that our selection criteria yield a galaxy population at $z\simeq 3.7\pm0.4$.

\section{The Physical Properties of Most UV-luminous LBGs}

\begin{figure}[t]
\epsscale{1.}
\plotone{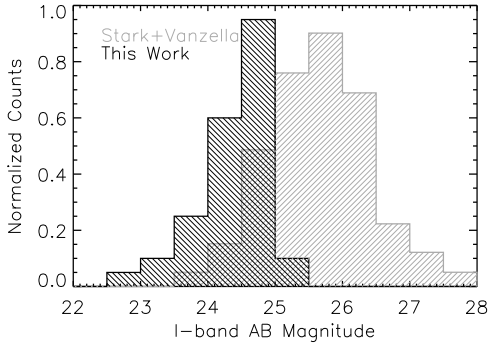}
\caption[compare_stark]
{
The normalized $I$-band magnitude distribution of our sample is compared to that of the Vanzella+Stark sample \citep{vanzella09, stark10}. The galaxies in our sample are roughly 2.1 times more UV-luminous than those in the Vanzella+Stark sample. 
}
\label{compare_stark}
\end{figure}

The spectra provide rich details about the physical properties of the most UV-luminous galaxies and show striking differences in the galaxy properties at different UV luminosities (i.e., star-formation rates). Before we proceed, it is useful to compare our sample to other large spectroscopic samples (generally of less UV-luminous galaxies) at the same redshift. \citet{vanzella06,vanzella09} observed $\approx50$ galaxies at $z\simeq3.7$ with the FORS2 spectrograph on the  Very Large Telescope. More recently, \citet{stark10} obtained a large sample ($\approx120$) of galaxies at $z=3.0-4.5$ with the DEIMOS on the Keck II Telescope. Both samples were drawn from the same photometric candidates selected from the Great Observatories Origins Deep Survey \citep{mauro04a}, using the Lyman break technique described in \S 2. In Figure \ref{compare_stark}, we compare the magnitude distribution of our sample to those of the combined Vanzella+Stark samples. The difference in the median luminosities between the two samples  is  0.8 mag. As the redshift selection functions for these samples are nearly identical, the median UV luminosity of our sample is roughly 2.1 times higher than that of the Vanzella+Stark samples. 

\subsection{The Ly$\alpha$ Equivalent Width Distribution}\label{Lya_EW}

We measure the Ly$\alpha$ EW for each galaxy directly from the one-dimensional spectra. The average continuum level $c_{\rm{red}}$ is determined by taking a weighted average of the pixels in the region corresponding to the rest-frame $1225-1255$\AA\ (and excluding all pixels affected by OH sky lines). The line flux  $F_{\rm{Ly\alpha}}$ is computed by summing the flux in excess of the average continuum level $c_{\rm{red}}$  at $\lambda_{\rm{rest}}=1213\rm{-}1221$\AA, and the rest-frame EW $W_0$ is then estimated as $W_0\equiv F_{\rm{Ly\alpha}}/ [c_{\rm{red}}(1+z)]\Delta \lambda$. Positive $W_0$ represents emission. Random errors in the EWs are estimated from the uncertainties in the line flux and continuum level. 

\begin{figure*}[t]
\epsscale{1.}
\plotone{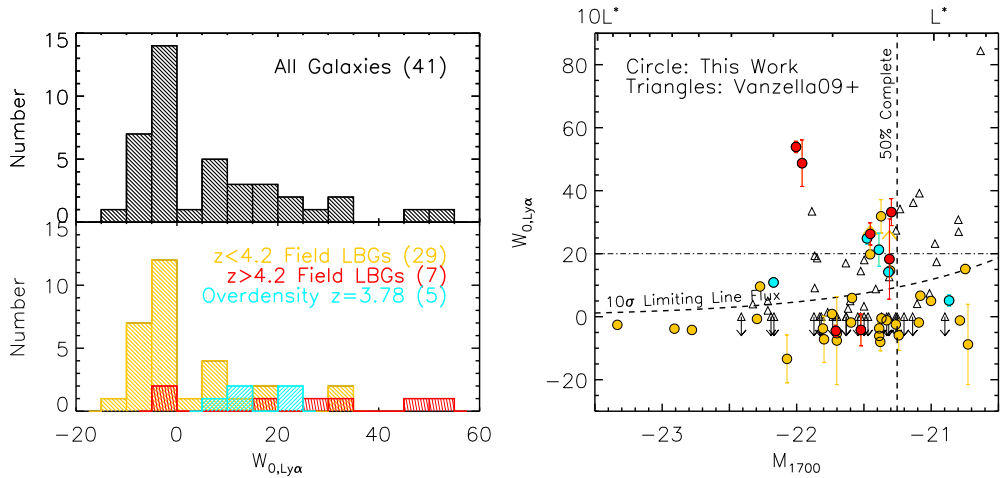}
\caption[fig_ew_muv]
{
{\it Left:} The distribution of the rest-frame Ly$\alpha$ EWs for the entire spectroscopic sample (top panel) and separated into the 5 galaxies in the $z=3.78$ structure and the 36 ``field'' galaxies (bottom panel). The difference in the median values ($W_0=14.2$\AA\ and $-1.2$\AA, respectively) between the two samples is significant. {\it Right:} The distribution of the Ly$\alpha$ EWs as a function of UV  luminosity suggests a lack of strong Ly$\alpha$ emitters at the very bright end. The majority of the high-EW sources are at the high-redshift tail of our selection function, i.e., at $z>4.2$, which is likely a result of the spectroscopic selection bias. On the other hand, the relatively strong Ly$\alpha$ emission from the $z=3.78$ structure appears to be real, with 4/5 sources observed with $W_0>10$\AA. When $z>4.2$ galaxies and those at $z=3.78$ are discarded, only 1/29 (i.e., 3\%) galaxies have $W_0\geq 20$\AA. 
}
\label{fig_ew_muv}
\end{figure*}

In Figure \ref{fig_ew_muv}, we show the distribution of the rest-frame Ly$\alpha$ EWs for all LBGs in our sample. Even though a half of our sample exhibit some level of Ly$\alpha$ in emission,  the median EW is negative. For the 36 ``field'' galaxies, the median $W_0$ is $-1.2$\AA. On the other hand, the 5 ``protocluster'' galaxies appear to have a much higher median EW of 14.2\AA. There are only 7/41 (17\%) galaxies that $W_0\geq20$\AA\ in our sample. Of those 7 galaxies, 4  are at $z>4.2$ and additional 2 belong to the $z=3.78$ structure (right panel of Figure \ref{fig_ew_muv}). If we accept the possibility that the majority of $z>4.2$ galaxies are selected because of their strong Ly$\alpha$ emission, and that the galaxies in the $z=3.78$ structure are not representative of the remaining ($z<4.2$) field sample, then only 1/32 ``field galaxies'' (3\%) have $W_0\geq20$\AA\ (BD235717, $W_0=32$\AA). These estimates are consistent with the measurements made by \citet[][see their Figure 13]{stark10} and extend these to higher UV luminosities.  

To investigate the frequency of high-EW sources with UV luminosity, we compute the luminosity of the sources at the rest-frame 1700\AA\ (Figure \ref{fig_ew_muv}, right). First, we adopted the \citet{madau95} prescription and calculated the effective flux attenuation by the intergalactic medium (IGM) in the $R$-band as a function of redshift, which changes from $\Delta m=0$ at $z<3.6$, to $\Delta m=0.17$ at $z=4$, to $\Delta m=0.39$ at $z=4.3$.  Second, from the IGM-corrected $(R-I)$ color, we estimate the UV spectral slope $\beta$ assuming a power-law spectral energy distribution  (i.e., $f_\lambda\propto \lambda^{\beta}$). The absolute magnitude at $\lambda=1700$\AA, $M_{1700}$, is computed by interpolating from the $I$-band total magnitude. For comparison, we also show the locations of the $B$ dropouts in the GOODS South \citep{vanzella09} as open triangles in the same figure. We computed the UV magnitudes at $1700$\AA, by interpolating from the $(i-z)$ color, both of which are free of intergalactic absorption at this redshift range. We note that \citet{vanzella09} did not measure the EWs for the galaxies with no Ly$\alpha$ emission, and therefore the triangle points at zero EW should be considered as an upper limit. 

To understand the spectroscopic bias, which preferentially selects strong Ly$\alpha$ line emitters, we computed the rest-frame EWs corresponding to the 10$\sigma$ limiting line flux ($\sim1.8\times 10^{-17}~\rm{erg}~\rm{cm}^{-2}~\rm{s}^{-1}$) at different continuum luminosities at rest-frame 1700\AA\ (dashed line). We assumed the spectral slope $\beta=-1.6$, consistent with the average value measured at this luminosity range \citep{lee11}. Also shown in Figure \ref{fig_ew_muv} (right panel) as vertical dashed line is the 50\% completeness limit in $I$-band source detection based on the image simulations we carried out on these fields. These considerations suggest that our data allow to identify line emission at the 10$\sigma$ level for galaxies with  $\rm{EW}\gtrsim10$\AA\ at the faint end of our sample. 

Due to the small sample size, it is difficult to draw a conclusion based on our sample alone. However, the picture becomes clearer when we consider the Vanzella sample and ours jointly. The observations of the former should be at least as sensitive as our observations, and therefore their 10$\sigma$ limit should be comparable to ours (dashed line in Figure \ref{fig_ew_muv} right).  From the combined sample, the increasing frequency of higher-EW sources (i.e., those with stronger Ly$\alpha$ emission) toward faint luminosities is evident. Interestingly, we find that there are no strong Ly$\alpha$ emitters (i.e., with $W_0>10$\AA) at $M_{\rm{UV}}<-22.2$ ($L_{\rm{UV}}\geq 3L^*_{z\simeq3.7}$) in the combined sample even though there is no selection bias against identifying such sources. The lack of very bright, high-EW sources at $z\sim4$ appears to be in conflict with that found at $z\sim6.0\rm{-}6.5$ \citep{curtis-lake12}.  While reversal of such a trend could be potentially interesting, better statistics are needed to place stronger constraints on the Ly$\alpha$-emitting fraction at the brightest end.

\subsection{Large-scale Outflows}\label{outflows}

Interstellar absorption lines that are blue-shifted with respect to the galaxy's systemic redshift are frequently observed in local and high-redshift star-forming galaxies over a wide range of luminosities and masses. These observations provide strong evidence of galactic-scale outflows that remove a considerable amount of neutral and ionized gas from the galaxy \citep{heckman00,martin05,martin06,schwartz04, shapley03, rupke05,adelberger05, vanzella09,weiner09,steidel10}. The quantity of material ejected from a galaxy into the IGM and the physical process(es) responsible are critical to our understanding of ``feedback", which is thought to shape a galaxy's star-formation, chemical and evolutionary history,  as well as the chemical history of the IGM. In particular, the relative efficiency of the feedback in galaxies of different luminosities/masses can shed light on their respective assembly histories. \citet{martin05} and \citet{weiner09} observed that the outflow velocity increases with the SFRs of the galaxy \citep[local ultra-luminous IR galaxies for the former, and $z\simeq1.4$ star-forming galaxies for the latter; also see][]{rupke05}. On the other hand, several more recent studies of local and intermediate-redshift galaxies do not find such a strong trend \citep{chen10,kornei12}, and the question regarding the luminosity dependence of feedback remains debated. 

At high redshift, measurements of outflows based on the largest sample of star-forming galaxies at $z\simeq2-3$ do not show any evidence of a luminosity dependence \citep{steidel10} and instead find marginal evidence for the opposite trend, i.e., that more massive galaxies have a lower outflow velocities as traced by the centroids of interstellar absorption lines \citep[see Figure 3 of][]{steidel10}. The discrepancy between studies at different redshifts is hard to understand, but may be in part explained by large scatter inherent to the measured outflow velocities, which sensitively depends on the viewing angle of the observer as well as the opening angle of the winds with respect to the galactic axis. Furthermore, different line diagnostics  studied at different redshifts may trace different kinematic components in the ISM, making direct comparison challenging. For example, the Na D doublet (arising in neutral gas) is used to trace the outflowing gas in local galaxies, whereas intermediate redshift studies use Fe~{\sc ii}, Mg~{\sc ii}, Mg~{\sc i}, which generally have higher ionization potentials. Furthermore, galaxies at lower redshift can be (and typically are) observed at a higher spectral resolution, where the systemic and outflowing components can be more easily separated \citep[e.g.,][]{chen10}; this approach is impractical for faint, high-redshift galaxies. Finally, the limited dynamic range in galaxy properties in most high-redshift samples studied to date may also play a role in erasing any trend of outflow velocity with other galaxy properties, especially given the inherently large scatter in the measurements or in the intrinsic distributions. Indeed, the local correlation is largely driven by the measurements of dwarf starbursts and (Ultra-)luminous IR galaxies \citep{schwartz04,martin05}. In this section, we attempt to improve on previous studies of outflows by extending the investigation to the more luminous high-redshift sample of galaxies uncovered by our survey.

\begin{figure*}
\epsscale{1.0}
\plottwo{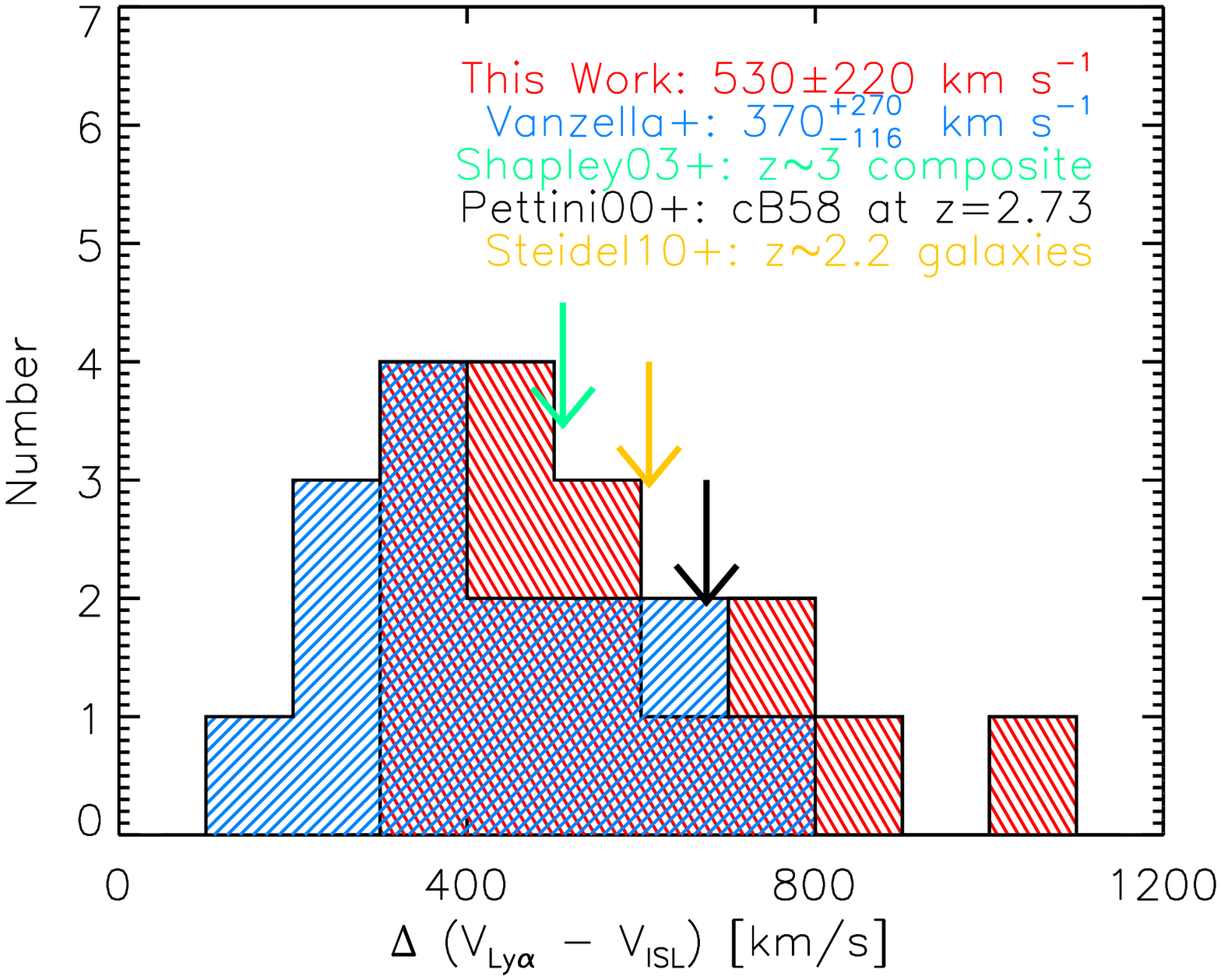}{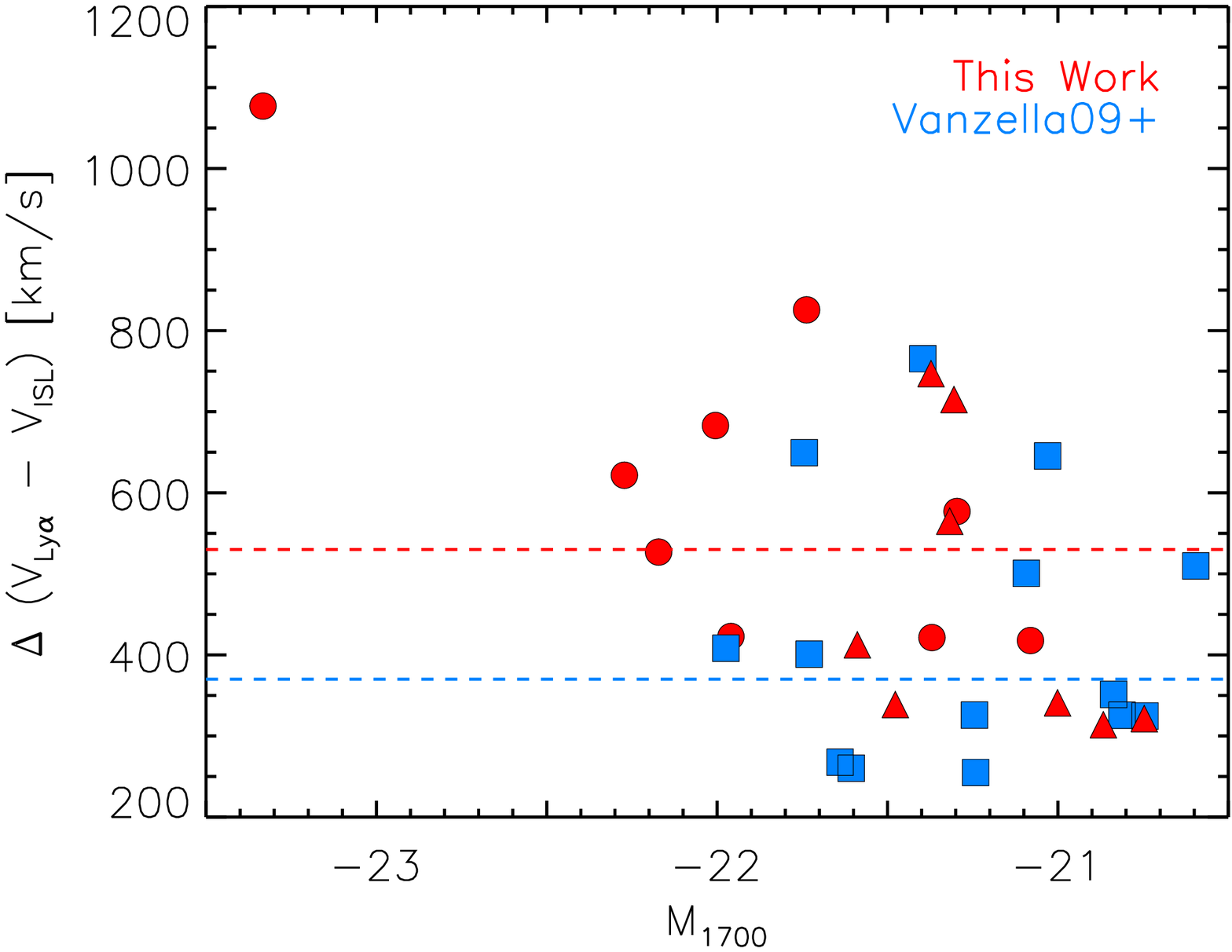}
\caption[fig_vdiff]
{
{\bf Left:}  The red histogram shows the distribution of the relative velocity between Ly$\alpha$ emission and interstellar absorption lines, measured from the 18 galaxies that clearly show both. The distribution of less UV-luminous $z\simeq 3.7$ galaxies in the GOODS South field measured by \citet{vanzella09} is shown in a blue hatched histogram. The median velocity of the galaxies in our sample is larger by $\approx 160$ km~s$^{-1}$ than the Vanzella sample. The same quantities measured for a large sample of $z\simeq 3$ galaxies \citep{shapley03} and that of MS 1512-cB58 \citep{pettini00} are also indicated. 
{\bf Right:} The correlation between the relative velocity inferred from the Ly$\alpha$ emission and interstellar absorption redshifts, $\Delta (v_{\rm{Ly}\alpha}-v_{\rm{abs}})$, and UV luminosity ($M_{1700}$) is shown together with the Vanzella sample (filled squares). The galaxies that are detected (undetected) in the IRAC data are shown in filled circles (triangles). 
}
\label{fig_vdiff}
\end{figure*}

The outflow velocity is, by definition, the velocity of the ISM material relative to the galaxy's systemic redshift. The latter is usually measured by either stellar absorption lines or non-resonant emission lines from ionized or neutral gas. Unfortunately, most nebular emission lines in the rest-frame UV (e.g., Ly$\alpha$ and \ion{C}{4}; covered by our observed optical spectra) arise from resonant transitions and are strongly affected by radiative transfer effects; the ones that are not resonant (e.g., \ion{He}{2}, \ion{C}{3}]) are generally too weak to have significant detections in the existing spectra. In addition, at the redshift of our sample the most prominent nebular lines (such as H$\alpha$, [O {\sc iii}]4959,5007, [O {\sc ii}]3727) are redshifted into the near-infrared (some beyond the $K$-band) at $z\gtrsim 3.8$.

Instead, we measure the relative velocity between the Ly$\alpha$ emission and that of the interstellar lines as a proxy for the outflowing medium. \citet{verhamme06} showed that the emergent Ly$\alpha$ profile observed at high redshift can be qualitatively reproduced by a simple model in which the energy from supernovae explosions drives an expanding shell of uniform velocity $V_{\rm{exp}}$. In their model, the Ly$\alpha$ photons that are back-scattered from the receding end of the expanding shell have a higher escape fraction by being scattered out of the core of the line; this results in redshifting the peak of the Ly$\alpha$ emission by a relative velocity of $\delta v=2V_{\rm{exp}}$. 
On the other hand, interstellar absorption in the advancing side (i.e., the observer's side) of the expanding medium causes a trough that is blueshifted by $\delta v=-V_{\rm{exp}}$. In reality, there is observational evidence that some of the details of the Ly$\alpha$ profile is inconsistent with such a simple model \citep[][]{quider09,kulas12} not to mention that there is evidence that ISM is neither uniform nor at a constant velocity \citep[e.g.,][]{steidel10}. Nevertheless, if both Ly$\alpha$ emission and interstellar absorption are primarily shaped by the outflowing medium, the relative velocity between the two should roughly trace the outflow velocity to first order. 

We measured the relative velocity $\Delta V \equiv V_{\rm Ly\alpha}-V_{\rm ISL}$ between the interstellar and Ly$\alpha$ features for the 17 galaxies with spectra of sufficient S/N  to detect both. The velocity ranges from 235~km~s$^{-1}$ to 1140~km~s$^{-1}$ as presented in Figure \ref{fig_vdiff}. The median (mean) value for the sample is 527 (531) ~km~s$^{-1}$ with the standard deviation of 220~km~s$^{-1}$. These values are significantly larger than the offset expected due to the intergalactic absorption alone\footnote{Assuming the \citet{madau95}  prescription and an initially Gaussian line profile, the IGM absorption causes  the centroid of Ly$\alpha$ line to shift by $-70$~km~s$^{-1}$ at the median redshift of $z=3.7$. The shift ranges over $-(50-85)$~km~s$^{-1}$ at the  redshift range of $z=3.3-4.1$.}, and thus strongly supports the presence of the ISM in motion. 
In Figure \ref{fig_vdiff} (left), we show the relative velocity distribution of the 17 galaxies in our sample, along with the measurements for the highly magnified $z=2.96$ galaxy MS 1512-cB58 and the average values for $z\simeq 3$ and $\simeq 2$ star-forming galaxies \citep[][respectively]{pettini00,shapley03,steidel10}. Also shown is the velocity distribution of 16  galaxies at $z\simeq 3.7$ measured by  \citet[][]{vanzella09}. The median velocity measured for the Vanzella sample is 370~km~s$^{-1}$, $\approx$160~km~s$^{-1}$ lower than the value for our sample, although both have a substantial scatter. To test the significance of the difference between the two distributions, we used the Kolmogorov-Smirnov test (K--S test, hereafter) in IDL {\tt kstwo.pro}.  The test returned the probability $P=0.286$ that the two measurements are drawn randomly from the same distribution. 

Using all 33 galaxies (from our study and the Vanzella et al. 2009 sample) with $\Delta V$ measurements, we investigate whether the outflow velocity varies with UV luminosity and stellar mass. Figure \ref{fig_vdiff} (right) shows the outflow measure $\Delta V$ as a function of UV luminosity, $M_{1700}$. The two horizontal lines mark the median value for the two samples. The data show a lot of scatter, as might be expected if orientation geometry, radiative transfer effects, or other physical processes play a role in determining the average $\Delta V$ of a galaxy. However, there is a suggestion that the ``upper envelope'' of the outflow velocity is larger for more luminous galaxies.  We used two statistics to test for the presence of a correlation: Kendall's $\tau$ and Spearman $\rho$ tests. Both methods measure the test statistics ($\rho_{\rm{SR}}$ or $\tau_{\rm{K}}$) and the probability of a null hypothesis (i.e., no correlation; $P_{\rm{SR}}$ or $P_{\rm{K}}$). The tests on our sample of 17 galaxies returned ($-$0.593, 0.012) and ($-$0.456,0.011) for the Spearman $\rho$ and Kendall $\tau$ test, respectively, and reject the null hypothesis (no correlation) at the $\approx99\%$ level.  Inclusion of the Vanzella sample results in measures of (-0.289,0.103) and (-0.216,0.077),  slightly weakening the strength of this statement, but still suggestive of a weak correlation between our outflow velocity measure $\Delta V$ and the UV luminosity of a galaxy. Because the most luminous galaxy in our sample (BD18449) also has the highest velocity offset (1140 km~s$^{-1}$), we repeated the same tests excluding the object. The probability of null hypothesis is 4\% ($P=0.04$) for our sample (16 galaxies), and 17\% for the combined sample (32 galaxies).

At $3.2<z<4.2$, the IRAC [3.6$\mu m$] band samples the rest-frame $I$-band, and thus can be used as a proxy for the stellar mass content. Unfortunately, the existing data from the Spitzer Deep Wide-Field Survey \citep[SDWFS;][]{ashby09} are too shallow to robustly detect $z\simeq 3.7$ galaxies with high S/N \citep[the $3\sigma$ depth at 3.6$\mu$m is 23.29~AB mag; see Table 5 of][]{ashby09} .  Of the 17 galaxies with outflow measurements, 6 are detected in the [3.6$\mu m$] SDWFS imaging, 3 are ambiguous, and the remaining 8 are not detected. In Figure \ref{fig_vdiff} (right), the IRAC-detected and IRAC-undetected sources are marked as red circles and triangles, respectively. The Vanzella sample is drawn from the GOODS-South survey field which has much deeper {\it Spitzer} coverage \citep[the 3$\sigma$ depth at 3.6$\mu$m of 26.75~AB; see Table~1 of][]{lee12a}. All 16 galaxies in the Vanzella sample are securely detected at [3.6~$\mu m$] with magnitudes ranging over $22.88-25.43$ AB mag with a median of 23.96 AB mag. Given the depth of the SDWFS data, only two of the galaxies in the Vanzella sample would have been detected had they been in the Bo\"otes field.

Within our sample, there is weak evidence of a correlation between outflow velocity and IRAC brightness. The median (mean) relative velocities are $\Delta V \equiv (V_{\rm{Ly}\alpha}-V_{\rm{abs}}) = 622 (655)\pm226$ and $412(481)\pm213$ km~s$^{-1}$ for the IRAC-detected and IRAC-undetected galaxies, respectively. Removing the outlying galaxy with the highest velocity, BD18449, from the IRAC-detected subset changes the median to $577 (566)\pm99$  km~s$^{-1}$. If we include the two galaxies that are marginally detected in IRAC, the relative velocity decreases slightly to $577(593)\pm219$ and $526(524)\pm108$ km~s$^{-1}$ with and without the source BD18449, respectively. In short, IRAC-bright galaxies appear to have a higher value of $\Delta V$ than their fainter counterparts. Because we do not have robust IRAC photometry for the majority of our galaxies, we are unable to directly test a bivariate correlation. A K--S test indicates that the distribution of the IRAC-undetected galaxies in our sample is statistically identical to that of the Vanzella sample. On the other hand, the IRAC-detected galaxies are likely drawn from a different distribution, with just 7.9\% probability of a null hypothesis when compared against the Vanzella sample. 

When we consider only the galaxies within the Vanzella sample, we find no correlation with luminosity, IRAC brightness, or stellar mass. The test statistics are inconclusive, and can only rule out the null hypothesis at the $\approx 45\%$ probability level.

It is not clear why our sample shows a correlation while the Vanzella sample does not. One possible explanation is that our sample probes a wider dynamic range of UV and optical rest-frame luminosities and thus is better suited to discerning a trend. The unknown viewing angle of our observations relative to the direction of the outflows will result in a large scatter in the observed velocities of outflows, which can wash away any trend if the dynamic range is not large enough.  While better sampling of galaxies at the most luminous or massive end is clearly needed to place the constraints on a firmer footing, our results are qualitatively consistent with 
studies of lower-redshift star-forming galaxies \citep{martin05,rupke05,weiner09}, where the outflow velocity increases with a galaxy's SFR and mass. 

\subsection{The Average Physical Properties of the Luminous $z\simeq 3.7$ Galaxies}\label{composite}

Creating a high S/N composite spectrum, by averaging over many individual spectra, can allow us to discern spectral features that are not typically visible in low-S/N individual spectra, and thereby infer the physical properties of the ``typical'' sample galaxy in a greater detail \citep[e.g.,][]{steidel01,shapley03, vanzella09}. The main challenge of constructing a composite spectrum is the fact that we do not have a good estimate of the systemic velocity of each galaxy. The spectral features we use for redshift identification, Ly$\alpha$ emission and/or interstellar absorption lines, are biased with respect to the systemic velocity. As discussed in \S\ref{outflows}, the interstellar absorption lines are often observed to be blue shifted with respect to the galaxy's systemic redshift, whereas the peak of the Ly$\alpha$ emission is redshifted due to radiative transfer effects. 

Deriving the systemic redshift using a statistical correction is difficult to justify. Most derived corrections are based on LBG samples at lower luminosity and / or lower redshift  \citep[e.g.,][]{adelberger03, steidel10}. In light of our findings in \S\ref{outflows} that the outflow velocity may depend on the galaxy's luminosity, the suggested values from those studies may not reasonably apply to our sample. Instead, we compute a fiducial redshift for each galaxy and use it to de-redshift each spectrum. For the galaxies with only interstellar absorption lines, which we refer to as ``abs''-type galaxies, the fiducial redshift is computed by applying a constant offset $v$ to the measured interstellar redshift;  i.e., $z_{\rm{fid}}=(-v/c+z_{\rm{IS}})/(1+v/c)$ where $c$ is the speed of light. We assumed the offset value of  $v=-250~\rm{km}~{s}^{-1}$, which places S~{\sc v} photospheric absorption line at the correct vacuum wavelength, i.e., $\lambda=1501.76$\AA. As for the galaxies with both Ly$\alpha$ and interstellar redshifts (``comp''-type), we take the mean of the two as fiducial redshift. This approach should minimize the effect from the range of outflow velocities observed in our sample on our composite spectrum, by making a larger correction if the outflow velocity of a given galaxy is larger. Furthermore, the mean correction applied to the ``comp'' galaxies should be  $v\simeq -250~\rm{km}~{s}^{-1}$ with respect to the interstellar redshifts (see Figure \ref{fig_vdiff}, the median velocity offset between Ly$\alpha$ and IS components is $530~ \rm{km}~{s}^{-1}$), similar to that applied to the ``abs'' galaxies. 

Because we base our stacking on the interstellar redshifts, we therefore limit our analyses only to the galaxies with measured interstellar redshifts. All but four galaxies in our sample have reliable interstellar redshifts determined from at least two absorption lines, while only half (20) have detectable Ly$\alpha$ emission. Another advantage of limiting the analyses to those with interstellar redshifts is that the composite spectrum created as a result should provide a fair representation of a flux-limited sample of high-redshift galaxies with no bias towards the Ly$\alpha$ line-emitting galaxies. 

We created the composite spectra as follows. First, we determine the fiducial redshift for each galaxy as described above. Second, we resample each spectrum to a common (near) rest-frame wavelength vector using cubic spline interpolation (IDL {\tt interpol.pro} with {\tt spline} option). Third, we normalize each spectrum using the average flux density at $1430-1470$\AA, where the spectrum is devoid of strong interstellar lines. Fourth, at each resampled wavelength element, we average the individual spectra weighting by the inverse variance \citep[where the inverse variance is derived from the $1\sigma$ error spectrum provided by the reduction pipeline; see][]{cooper12}. For each individual spectrum, the regions contaminated by atmospheric $A$ and $B$ band absorption (at $\lambda=7592-7675$\AA\ and $6886-6881$\AA, respectively) are masked and excluded from the average. Since the error spectrum is background-limited, each galaxy has roughly equal weight in the average regardless of its luminosity. 

\begin{figure*}[t]
\epsscale{1.}
\plotone{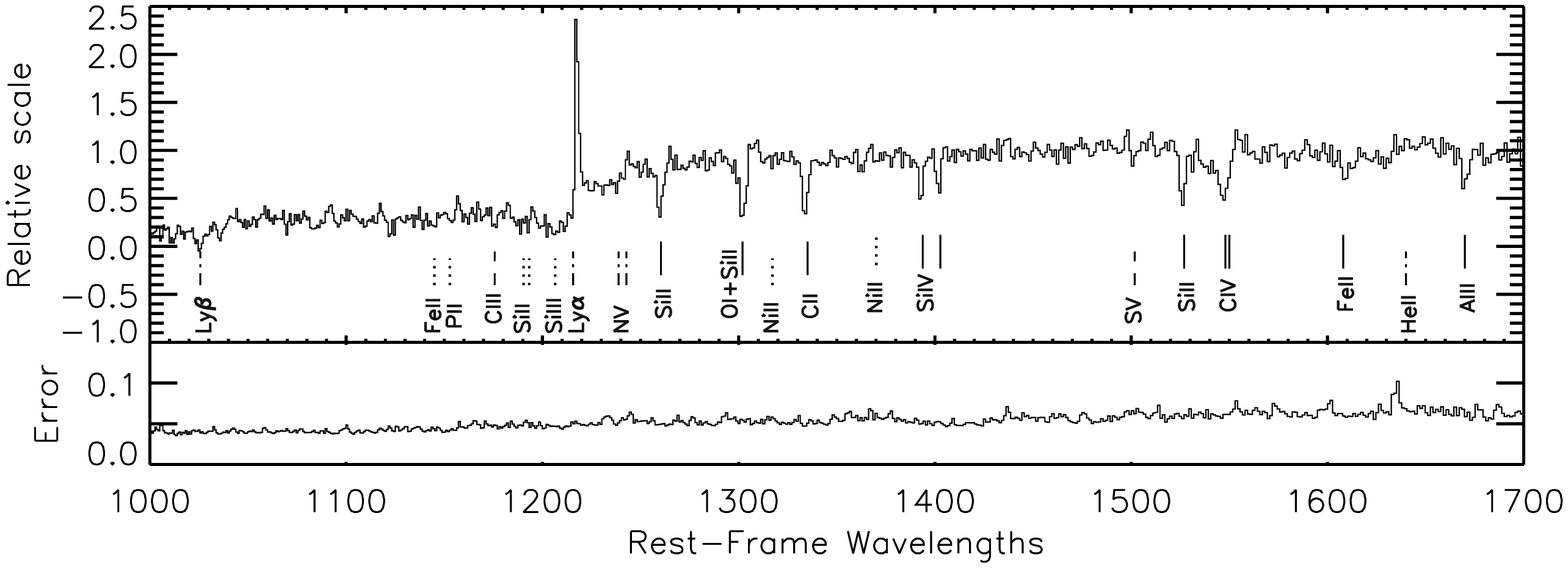}
\caption[coadd_all]
{
The average spectrum of all 36 confirmed star-forming galaxies at $3.2<z<4.6$ together with the error spectrum on bottom. The spectrum is normalized at $1430-1470$\AA. Also marked are the vacuum wavelengths of prominent stellar photospheric (dashed), nebular (dashed-dot), and interstellar lines (strong lines in solid, weaker lines in dotted lines, respectively). Note that while the expected line centroids are labeled, not all these features are detected in the composite spectrum. 
}
\label{coadd_all}
\end{figure*}

In Figure \ref{coadd_all}, we show the composite spectrum constructed from 36 galaxies in our sample; i.e., all excluding five galaxies with no interstellar redshift. In the spectrum, we detect C~{\sc iii}~$\lambda$1176 line and S~{\sc v}~$\lambda$1502, which arise from stellar photospheric absorption. The lines are at the correct wavelengths, therefore confirming that our final spectrum is in fact at systemic. Finally,  we note that the bump in the error spectrum at $\lambda_{\rm{rest}}\approx1630$\AA\ shown in Figure \ref{coadd_all} (also in Figure \ref{coadd_abs}) is a result of the atmospheric $A$ band entering into the region for many  galaxies in our sample, and is not related to the He~{\sc ii} emission. 

\subsubsection{The Galaxies with no Ly$\alpha$ emission}\label{Lya_nonemitters}
\begin{figure*}[t]
\epsscale{1.}
\plotone{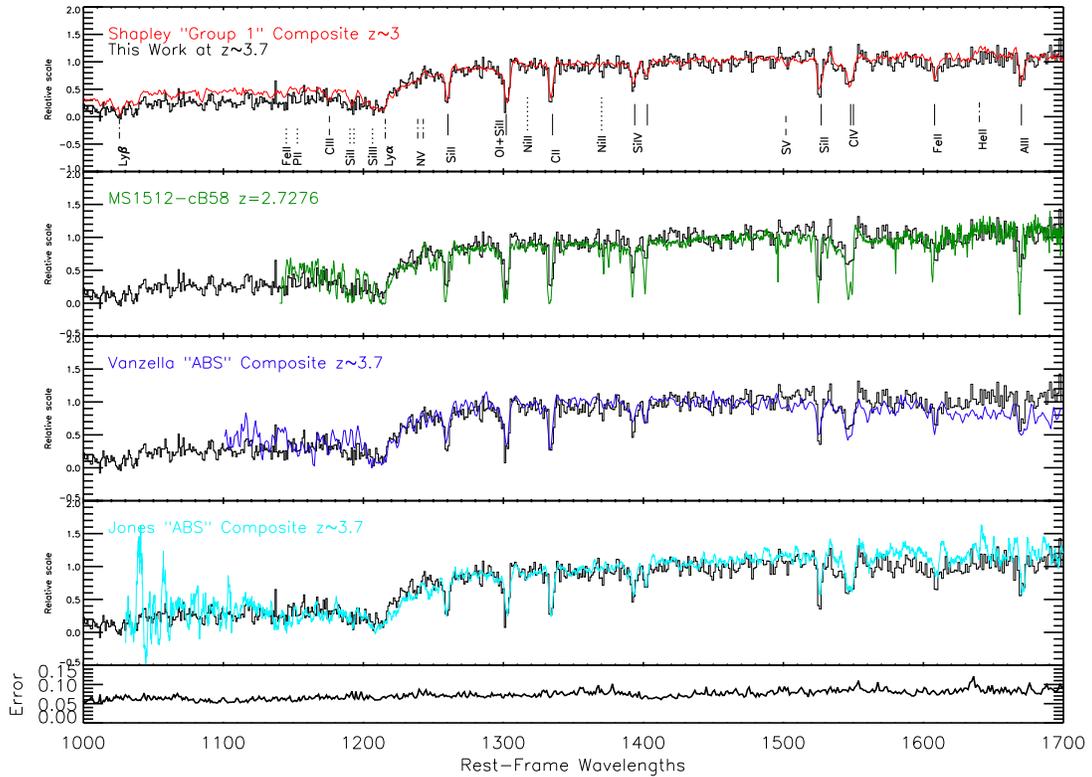}
\caption[coadd_abs]
{
The top panel shows average spectrum of galaxies without Ly$\alpha$ emission in our sample (black)compared with the  composite spectrum of the $z\simeq 3$ galaxies from \citet{shapley03} (red). The Shapley et al. spectrum shown here is their ``Group 1" composite which averages spectra of galaxies with Ly$\alpha$ EW in the bottom quartile of the overall distribution \citep[see][for details]{shapley03}. The lower panels show the same average spectrum of our sample compared with the spectra of MS1512-cB58 \citep[second panel][]{pettini02};  the $z\simeq 3.7$ composite of lower-luminosity galaxies \citep[third and fourth panel, see][respectively]{vanzella09, jones12}. Both composites are created from galaxies with no Ly$\alpha$ emission. 
}
\label{coadd_abs}
\end{figure*}

In Figure \ref{coadd_abs}, we show the composite spectrum for our $z\simeq  3.7$ sample constructed from 20 sources with interstellar absorption line redshifts and no Ly$\alpha$ emission. The spectrum clearly shows not only strong interstellar lines such as \si1260, O~{\sc i}+Si~{\sc ii} $\lambda$1303, \c1334, Si~{\sc iv} $\lambda\lambda$1393,1402, Si~{\sc ii} $\lambda$1526, C~{\sc iv} $\lambda\lambda$1548,1550, Fe~{\sc ii} $\lambda$1608, and Al~{\sc ii} $\lambda$1670, but also weaker features such as Ni~{\sc ii} $\lambda$1370,  Ly$\beta$, Si~{\sc iii} $\lambda$1206, N~{\sc v} $\lambda$~1239,1243.

The top panel of Figure \ref{coadd_abs} compares our composite spectrum with that of the average $z\simeq 3$ LBG from \citet{shapley03}. Specifically, the \citet{shapley03} spectrum shown is the composite constructed by averaging the 199 galaxies in the bottom quartile of the Ly$\alpha$ EW distribution (i.e., referred to as ``Group 1" by Shapley et al.; see their Table 3) and exhibits the strongest interstellar absorption. At $\lambda_{\rm{rest}}>1216$\AA, the $z\simeq 3$ and $z\simeq 3.7$ composite spectra closely mirror each other, even down to the small scale wiggles which likely result from weak nebular/interstellar features. The Ly$\alpha$ line morphologies blue-ward of the trough are also similar in the two spectra, suggesting that the column density and velocity distribution of the absorbing gas are comparable. While the interstellar line ratios in our sample are generally similar to those of $z\simeq  3$ galaxies, the EWs appear to be slightly larger in our spectrum (see \S\ref{interstellar_ew} later). 

\begin{figure*}
\epsscale{1.}
\plotone{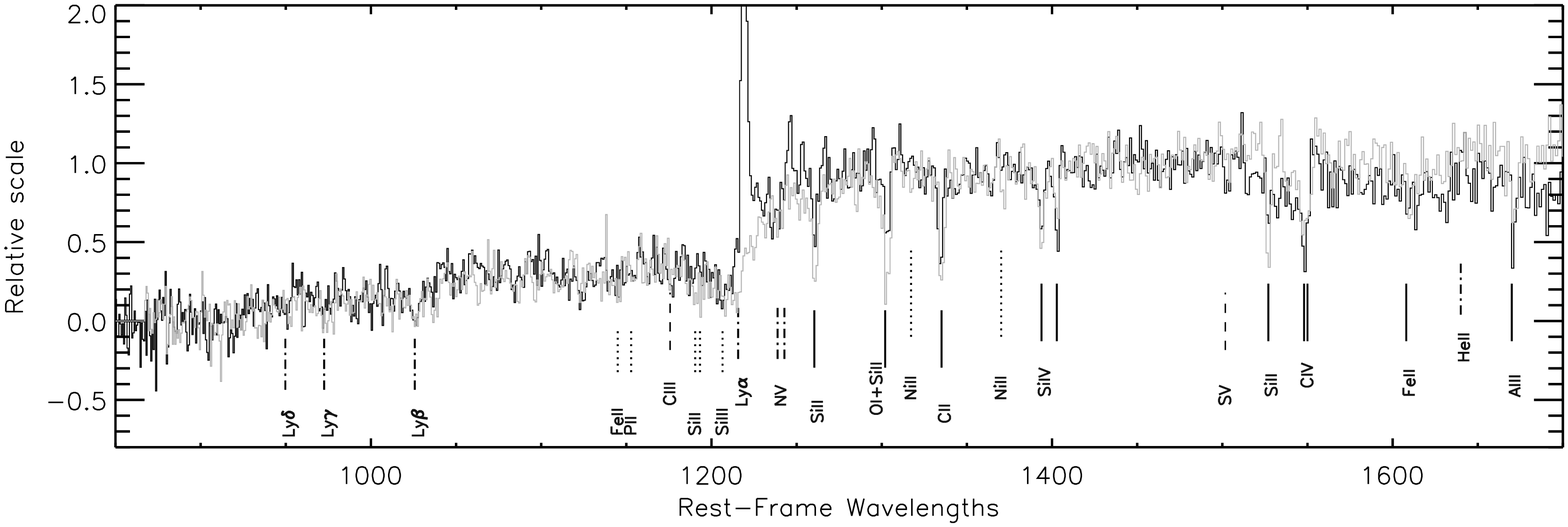}
\caption[coadd_em]
{
The composite spectra of the Ly$\alpha$-emitting galaxies (black) vs  non-emitters (light grey) are shown. Both spectra are normalized at 1450\AA. Interstellar absorption is generally stronger in the non-emitter spectrum, consistent with the expectation of a higher gas column density in the ISM for the non-emitters. We also detect Ni~{\sc ii}  in the non-emitter spectrum but not in the emitter spectrum. Conversely, S~{\sc v} is visible in the emitter spectrum, but not in the non-emitter spectrum (see discussions in \S\ref{Lya_emitters}).
}
\label{coadd_em}
\end{figure*}

Despite the general similarities of our $z\simeq 3.7$ sample to the $z\simeq 3$ galaxies, there  are also  several features that may be distinct.  Most notably, The C~{\sc iv} P Cygni emission, which traces stellar winds from massive stars and therefore sensitive to the stellar initial mass function and/or population ages \citep[see][and references therein]{pettini00}, may be enhanced for the $z\simeq 3.7$ luminous LBGs. Both spectra (top panel; Figure \ref{coadd_abs}) are normalized at 1450\AA\ and the continuum level at 1550\AA\ immediately following the C~{\sc iv} emission is not well determined as can be seen in Figure \ref{coadd_abs}. However, adjusting the normalization to a higher value would make the P Cygni emission more dramatically different. Although the difference is not statistically significant (as can be seen from the error spectrum on bottom of Figure \ref{coadd_abs}),  the composite spectra of lower-luminosity galaxies at $z\simeq 3.7$ presented in \citet[][third row]{vanzella09} and \citet[][fourth row in Figure \ref{coadd_abs}]{jones12} also appear to exhibit similarly strong C~{\sc iv} emission, lending support to the possibility that C~{\sc iv} emission may have been indeed stronger at higher redshift. 

We also detect N~{\sc v} $\lambda1240$ P Cygni feature at a lower significance. The absorption of the N~{\sc v} doublet is not seen in the $z\simeq 3$ Shapley composite, but observed in the UV spectrum of MS1512-cB58 to have $W_0=0.23$\AA\ \citep[second panel of Figure \ref{coadd_abs}:][]{pettini02}. In fact, the line morphologies of both C~{\sc iv} and N~{\sc v} in our spectrum are more consistent with that of cB58 than with the $z\simeq 3$ composite, even though cB58 is observed with a much higher covering fraction (the residual intensity in the line cores is nearly zero). Like C~{\sc iv},  N~{\sc v} also traces stellar winds from massive O stars and is thus also sensitive to the galaxy's IMF, age, and metallicity.

We also detect possible weak absorption lines of Ni~{\sc ii} $\lambda$1370, and  Ni~{\sc ii} $\lambda$1317, which  are absent in the \citet{shapley03} spectrum but are detected in the UV spectrum of MS1512-cB58 at a $W_0\simeq 0.2$\AA\  level \citep{pettini02}. Ni is one of the Fe-peak elements released by Type Ia supernovae (SNe) and is found to be underabundant  in the cB58 spectrum by a factor of three relative to, e.g., Si. \citet{pettini02} argue that the observed underabundance of the Fe-peak elements in cB58 is  evidence of a young ($\sim300$~Myr) population age, since metal enrichment by longer-lived, intermediate-mass stars has not yet caught up with the elements released by short-lived OB stars. The tentative detection of Ni in our luminous $z \simeq 3.7$ galaxies may suggest that these are ``older'' (i.e., have larger UV-luminosity weighted ages) than the lower-luminosity $z\simeq 3$ galaxies, perhaps because the more luminous galaxies have a more extended star-forming phase. Interestingly, population ages inferred from  photometric measurements of the Balmer break strength do no vary significantly with UV luminosities \citep[e.g.,][]{lee11,oesch12}, thus suggesting otherwise.  However, such measurements are almost always made on photometric samples, and typically show large scatter and uncertainty. While the current data have insufficient S/N to infer galaxy ages from absorption lines properties, future observations of deep, high-resolution spectroscopy of UV-luminous galaxies should provide independent estimates on galaxy's population age (as well as metallicity, and IMF). 

We compare our composite spectrum with the \citet{vanzella09} ``abs''-class composite created from 21 less-luminous $z\simeq 3.7$ galaxies which lack Ly$\alpha$ emission (third panel of Figure \ref{coadd_abs}). The absorption features are clearly stronger in our spectrum in most cases. The difference appears to be mainly luminosity-dependent as both composites have similar Ly$\alpha$ absorption EW. Our $z\simeq 3.7$ composite also has a redder UV slope than the Vanzella composite, consistent with the well-known correlation between luminosity and UV colors \citep[e.g.,][]{bouwens09, finkelstein12}. As in the case of the Shapley composite spectrum, Ni is not detected in the Vanzella spectrum. 

Finally, the bottom panel shows the ``abs''-type galaxy composite from the data presented in \citet{jones12}. The median luminosity of 24 galaxies included in the spectrum is $M_{\rm{UV}}\sim -21$. Their spectrum has a somewhat redder slope compared to our composite and the Vanzella composite, which may be due to systemic error in the flux calibration (T. Jones, private communication). The most notable differences are that the Jones et al. spectrum shows higher EWs in Si$^*$~{\sc ii} emission (at $\lambda=1265, 1309, 1533$\AA, but most pronounced at 1533\AA) and He~{\sc ii} emission at 1640\AA. We discuss this in further detail in \S\ref{emission_lines}. Similar to the Vanzella composite, the interstellar absorption lines in the Jones composite are generally weaker compared to our spectrum. We compare the interstellar absorption lines from these spectra in further detail  in \S\ref{interstellar_ew}.  

\subsubsection{Ly$\alpha$-emitting Galaxies versus Non-Ly$\alpha$ Emitters}\label{Lya_emitters}

The origin of Ly$\alpha$ emission in high-redshift galaxies is still not well understood. The observed Ly$\alpha$ EW is a sensitive function of the galaxy's age, metallicity, and the amount and relative geometry of gas and dust in the ISM. In theory, comparing the physical properties of galaxies with or without Ly$\alpha$ emission can provide useful information about which parameter is the driving factor of the observed Ly$\alpha$ emission in galaxies. Recent studies suggest that Ly$\alpha$-emitting galaxies are less dusty \citep{gawiser06,pentericci07,kornei10} on average than the non-emitters, although it is possible for very dusty galaxies to exhibit strong Ly$\alpha$ emission \citep[e.g.,][and the references therein]{finkelstein09,finkelstein11b}. Furthermore, different studies have conflicting results about how the Ly$\alpha$ emission is related to the galaxy's population age \citep{pentericci09,kornei10}. 

One complication that has hindered obtaining a clear picture has been that the galaxies observed with Ly$\alpha$ emission tend to be intrinsically less-luminous in their continuum emission. Combined with the spectroscopic selection effect (i.e., that it is easier to spectroscopically confirm galaxies with strong Ly$\alpha$ emission than those without), it is challenging to compare fairly the physical properties of the emitter and non-emitter populations. 

We directly compare the spectroscopic properties of the galaxies observed with and without Ly$\alpha$ emission by stacking them separately. Once again, we only consider the 36 (of the 41) galaxies with the measured interstellar redshifts. Of these, 17 galaxies exhibit Ly$\alpha$ emission and 20 do not. One galaxy in the Ly$\alpha$-emitting subsample has unusually strong N~{\sc v} emission, and we exclude this from the stack. We stacked the remaining 16 Ly$\alpha$-emitters and 20 non-emitters as described previously. The spectrum of non-emitters is identical to that shown in Figure \ref{coadd_abs}. 

Figure \ref{coadd_em} compares the two composite spectra (normalized at 1450\AA). The median (mean) UV luminosities of the two samples are comparable: $M_{1700}$ is -21.37 (-21.56) and -21.52 (-21.62) for the emitter and non-emitter subsamples, respectively. Similarly, the median (mean) $I$-band magnitudes are also comparable: 24.58 (24.40) and 24.41 (24.28), respectively.

It is clear from the spectra that most of the strongest interstellar absorption lines, including Si~{\sc ii}, O~{\sc i}+Si~{\sc ii}, C~{\sc ii} and Si~{\sc iv}$\lambda1394$, are stronger in the non-emitter composite spectrum. The anti-correlation between the EWs of Ly$\alpha$ emission and interstellar absorption was also observed in the $z\simeq  3$ galaxies by \citet{shapley03}. Shapley et al. argued it as evidence that the emergent Ly$\alpha$ strength and interstellar absorption features are jointly determined by random sight lines seen through the regions of differing optical depths: more optically thick sightlines result in more absorption/attenuation of Ly$\alpha$ and stronger interstellar absorption.

We also observe several subtle but intriguing features that have not been previously reported. The two Ni~{\sc ii} lines detected in the non-emitter spectrum are absent in the emitter spectrum. Although both Ni lines are detected at a low S/N level, the fact that both  appear in one spectrum and neither does in the other suggests that the trend is in fact real. In addition, S~{\sc v} line is clearly detected in the emitter spectrum but not in the non-emitter spectrum. Ni is one of the Fe-peak elements (Ni, Fe, Mn) that are released by Type Ia SNe, while Type II SNe are the main producers of S (together with Si, O, Mg, P). Another Type II element, N~{\sc v},   is  visible in the non-emitter population only. In the context of standard chemical evolution models, these observations provide unique insights into the recent star-formation history of the dominant stellar population. Based on these findings, we speculate that the luminous non-emitter population may have continued star-formation for a considerably longer period of time than the Ly$\alpha$-emitting population, as the latter have not had enough time to return metals produced by intermediate-mass, longer-lived stars into the ISM \citep[][]{pettini02}.  S~{\sc v} is not detected in either \citet{jones12} or \citet{vanzella09} emitter spectrum, suggesting that the process responsible for this line may also depend on luminosity (or, more likely, age).

\subsection{Emission Line Properties}\label{emission_lines}

\begin{figure*}[t]
\epsscale{1.}
\plotone{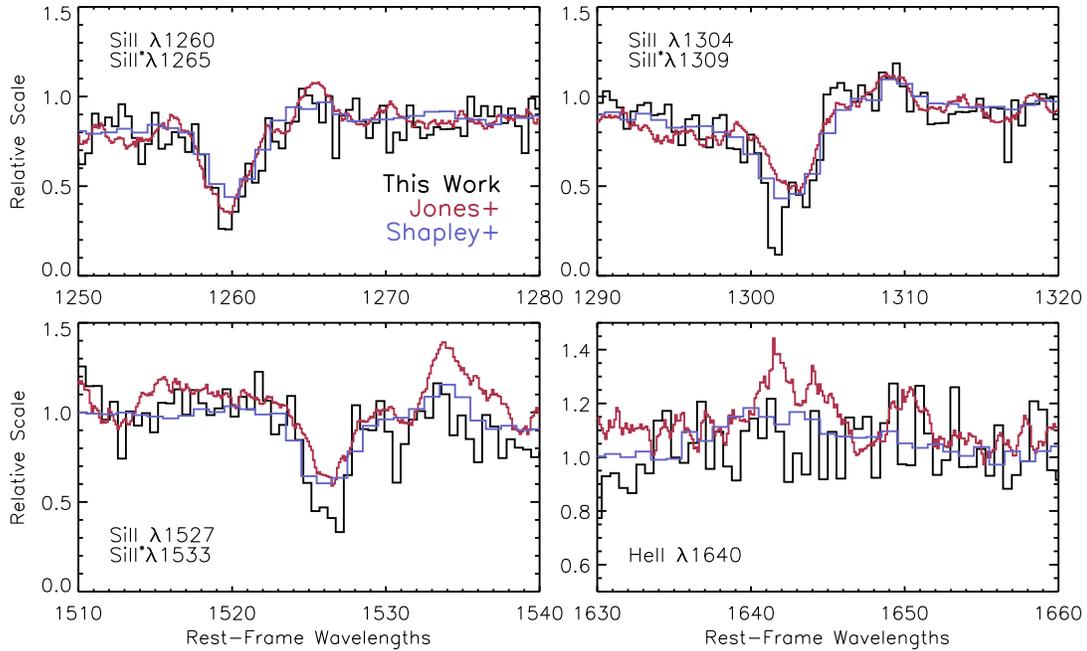}
\caption[compare_Si]
{
The composite spectrum showing the regions around Si~{\sc ii}$^*$ emission lines and He~{\sc ii} is compared with composite spectra of $z\simeq 3$ and $z\simeq 4$ LBGs from \citet{shapley03} and \citet{jones12} respectively. The Si~{\sc ii}$^*$ emission line profiles are very similar to those of the $z\simeq 3$ Shapley et al. (2003) composite, while the \citet{jones12} spectrum shows more pronounced emission. We do not detect He~{\sc ii} emission in our composite spectrum, although within margin of error ($\simeq 15$\% of continuum level), our spectrum is consistent with the $z\simeq 3$ Shapley et al. spectrum. }
\label{compare_Si}
\end{figure*}

We detect several weak Si emission line features in the composite spectrum. Si~{\sc ii}$^*$ emission lines ($\lambda=1265, 1309, \rm{and}~1533$\AA) are thought to arise from the fine-structure transitions within the outflowing gas in the ISMand CGM \citep[][]{shapley03}. Photons absorbed by a ground-state Si~{\sc ii} ion  is re-emitted  either at the same energy level or at the corresponding fine-structure transition with roughly equal probabilities. In Figure \ref{compare_Si}, we compare our composite spectrum (shown in Figure~\ref{coadd_all}) with those presented in \citet{shapley03} and \citet{jones12} at $z\simeq 3$ and $\simeq 4$, respectively. The Si~{\sc ii}$^*$ emission profiles in our spectrum are similar to those in the Shapley spectrum, while the \citet{jones12} spectrum shows clear excess Si~{\sc ii}$^*$ emission, most pronounced at 1533\AA. 

In the case of optically thick gas, resonant  Si~{\sc ii} photons will be continuously scattered until they emerge as Si~{\sc ii}$^*$, and therefore, the net EW of the Si~{\sc ii} transitions should be zero; i.e., $W_{\text{Si{\sc ii}*,em}} + W_{\text{Si{\sc ii},abs}}=0$. Contrary to this expectation, \citet{jones12} reported that the measured value for their $z\simeq 4$ sample is $W_{\text{Si{\sc ii}*,em}}/W_{\text{Si{\sc ii},abs}}=-0.53\pm 0.17$ when averaged over the two strongest Si~{\sc ii} transitions at 1260 and 1527\AA. In comparison, the analogous value measured from the Shapley composite at $z\simeq 3$ is $-0.16\pm0.04$. \citet{jones12} argued that the lower-than-expected strength of fine-structure emission is likely due to the fact that much of the emission arises in the CGM and therefore is not sampled within their 1\arcsec\ slitlets ($\sim 7$~kpc). They further argued that the radius of fine-structure emission may increase with redshift,  thus giving rise to a higher line ratio  $W_{\text{Si{\sc ii}*,em}}/W_{\text{Si{\sc ii},abs}}$ at higher  redshift as mentioned above. However, our DEIMOS observations cast doubt on this interpretation. Even though we used larger slit widths (1.2\arcsec\ sampling 8.6~kpc at $z=3.7$) than the Jones et al observations, the emission level is still comparable to that of the Shapley spectrum. One possible explanation is that the size of the Si~{\sc ii}$^*$-emitting CGM region is a strongly increasing function of galaxy luminosity rather than of redshift as \citet{jones12} proposed. 
 
Another notable emission line in the rest-frame ultraviolet spectrum is  He~{\sc ii} emission, which is clearly detected in the \citet{shapley03} and \citet{jones12} samples.  The line profile of the He~{\sc ii} emission (FWHM $\sim$  1500 km~s$^{-1}$ at $z\simeq 3$) suggests that the emission has been broadened significantly by fast, dense  winds from Wolf-Rayet (W-R) stars \citep{shapley03}. W-R stars, as  descendants of O stars with masses $\gtrsim 20-30M_\odot$, are short-lived, and thus the He~{\sc ii} emission of W-R origin is strong only in the initial phase of star-formation \citep{schaerer98}. Assuming constant star-formation history and  solar metallicity, \citet{schaerer98} estimated that EW(He~{\sc ii}1640) should drop to zero within $\approx 20$ Myr. Assuming subsolar metallicity ($0.4Z_\odot$), the He~{\sc ii}-bright phase decreases to just a few Myr.  

Our data do not show a clear sign of He~{\sc ii} emission (bottom right of Figure \ref{compare_Si}; also see Table \ref{tbl_3}) even though our spectrum is in reasonable agreement with the Shapley composite within (large) errors. We speculate that the lack of strong He~{\sc ii} emission may be due to the fact that the galaxies in our sample may have more extended star-formation histories than those in the Shapley and Jones sample. While a higher-resolution, higher S/N spectrum is necessary to place more robust constraints on the strength of the He~{\sc ii} emission in the luminous galaxies, such an interpretation is qualitatively in line with the inferences from other observations discussed earlier (\S\ref{Lya_nonemitters} and  \S\ref{Lya_emitters}) that the luminosity-weighted ages increase with UV luminosity and/or decrease with Ly$\alpha$ EW.

\subsection{Dependence of Interstellar Absorption Line Strengths on Galaxy Parameters}\label{interstellar_ew}
The strength of interstellar absorption in the UV spectrum is primarily determined by the kinematics, ionic column density, and covering fraction of the absorbing gas in the ISM, and therefore provides direct probes of the physical conditions therein. The fact that the interstellar absorption EW correlates so strongly with Ly$\alpha$ EW suggests the same gas that produces absorption features is also responsible for attenuating the Ly$\alpha$ emission \citep[e.g.,][]{shapley03,vanzella09}. When the line is saturated, which may be the case for many of the strongest interstellar absorption features at high redshift, the EWs primarily depend on the velocity dispersion of the gas (measured by deconvolved full-width-at-half-maximum) and covering fraction ($C_f$), and to a lesser degree, on the ionic column density.\footnote{The equivalent width of an unsaturated absorption line should obey $W_{\rm{IS}}\propto N_{\rm{ion}}$; for a saturated line, $W_{\rm{IS}}\propto b[\ln(N_{\rm{ion}}/b)]^{0.5}$ while $f_{\rm{max}}/f_{\rm{cont}}=1-C_f$ where $f_{\rm{max}}$ is the intrinsic flux density at maximum absorption, $b$ is Doppler parameter, and $C_f$ is covering fraction of the absorbing gas.} Hence, the measurements of such a correlation at different galaxy properties (e.g., luminosity, mass, color, morphology) and at different redshifts can provide valuable insights into the changing physical parameters in the ISM among the galaxies.  Towards that end, we compare the measurements of  interstellar lines with those in the literature. We mainly focus on the four strongest low-ionization interstellar (LIS) lines, Si~{\sc ii}~$\lambda$1260, O~{\sc i}+Si~{\sc ii}~$\lambda$1303, C~{\sc ii}~$\lambda$1334, and Si~{\sc ii}~$\lambda$1527, which trace cold neutral gas in the ISM. Following the convention used in \citet{shapley03} and \citet{jones12}, we use the average of these four as the indicator of LIS  line strength, and refer to it as $W_{\rm{LIS}}$ hereafter. 

\begin{deluxetable*}{ccccccc}
\tabletypesize{\scriptsize}
\tablewidth{0pc}
\tablecaption{Equivalent Widths Measured from the Composite Spectrum}
\tablehead{
\colhead{Ion} &
\colhead{$\lambda_{\rm{rest}}$\tablenotemark{a} (\AA)} &
\colhead{$W_{0,\rm{full}}$ (\AA)} &
\colhead{$W_{0,\rm{emitters}}$ (\AA)} &
\colhead{$W_{0,\rm{non-emitters}}$ (\AA)} &
\colhead{$W_{0,M_{\rm{UV}}<-21.4}$\tablenotemark{b} (\AA)} &
\colhead{$W_{0,M_{\rm{UV}}\geq-21.4}$\tablenotemark{b} (\AA)}}
\startdata
H~{\sc i} & 1215.67 & $4.5\pm0.2$& $15.8\pm0.1$& $-4.3\pm0.5$ & $-4.9\pm0.2$ & $-3.2\pm0.4$\\
Si~{\sc ii} & 1260.42 &$-1.8\pm0.2$  & $-1.4\pm0.2$ & $-2.3\pm0.3$ & $-3.0\pm1.0$ & $-2.0\pm0.5$\\
O~{\sc i}+Si~{\sc ii}& 1303.27 &$-2.7\pm0.2$  & $-1.9\pm0.2$& $-3.4\pm0.5$ & $-3.0\pm1.0$ & $-3.1\pm0.6$\\
C~{\sc ii} & 1334.53 &$-2.3\pm0.2$& $-1.6\pm0.2$& $-2.7\pm 0.3$& $-2.0\pm0.4$ & $-2.2\pm0.4$\\
Si~{\sc iv} & 1393.76 &$-1.4\pm0.1$& $-0.9\pm0.1$& $-1.8\pm0.2$& $-1.9\pm0.4$ & $-1.9\pm0.3$\\
Si~{\sc iv} & 1402.77 &$-1.0\pm0.1$& $-1.2\pm0.1$& $-0.9\pm0.1$ & $-0.7\pm0.3$ & $-1.1\pm0.2$\\
Si~{\sc ii} & 1526.71 &$-1.7\pm0.2$& $-0.8\pm0.2$& $-2.3\pm0.3$ & $-2.4\pm0.6$& $-2.1\pm0.3$\\
C~{\sc iv} & 1549.48 &$-3.1\pm0.3$& $-2.8\pm0.3$& $-3.3\pm0.3$ & $-3.4\pm0.6$ & $-3.1\pm0.4$\\
He~{\sc ii} & 1640.40 &$0.3\pm0.3$& $-0.2\pm0.5$&$0.1\pm0.5$ & $0.3\pm0.6$&$-1.1\pm1.0$
\enddata
\label{tbl_3}
\tablenotetext{a}{The vacuum wavelengths}
\tablenotetext{b}{The sample only includes the non-emitters in the sample (i.e., $W_{0,\rm{Ly}\alpha}<0$ \AA)}
\end{deluxetable*}

\begin{figure}[t]
\epsscale{1.}
\plotone{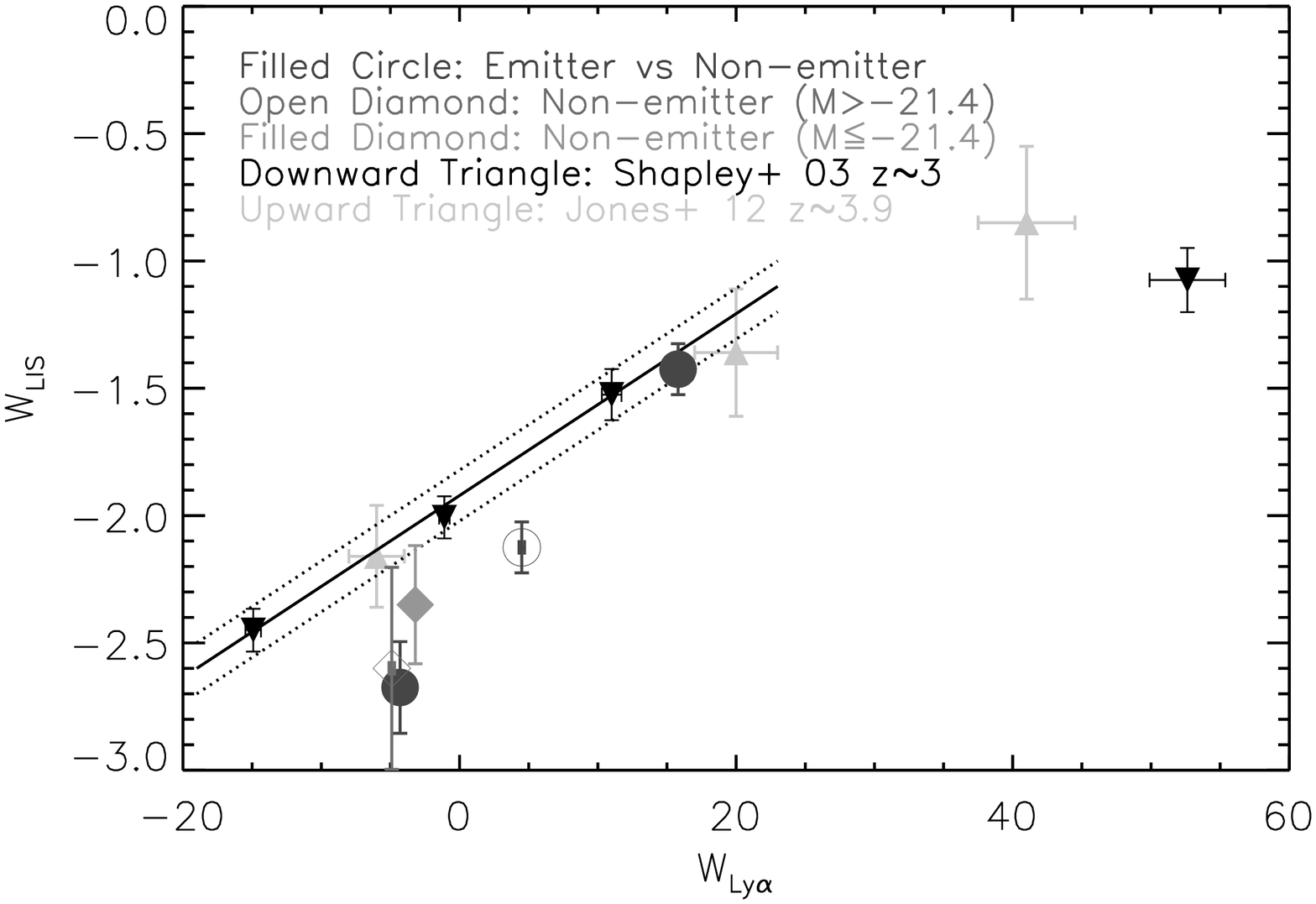}
\caption[fig_ew_los]
{
Equivalent width of low-ionization absorption lines as a function of Ly$\alpha$ EW. Large filled circles are from our sample divided into two bins as Ly$\alpha$ ``emitters'' and ``non-emitters'', while the large open circle shows the value measured for the full sample. Filled upward-pointing triangles show the measurements from the \citet{jones12} $z\simeq 3.9$ samples in three bins according to Ly$\alpha$ EW, while downward-pointing triangles indicate similar measurements at $z\simeq 3$ as published in \citet{shapley03}. Also shown are two luminosity bins ($M<-21.4$ and $M\geq -21.4$) of non-emitters in filled and open diamond, respectively. 
}
\label{fig_ew_los}
\end{figure}

We measured the $W_{\rm{LIS}}$ from five distinct composite spectra: (1) the full sample (Figure \ref{coadd_all}), (2,3) Ly$\alpha$-emitting galaxies and non-emitters (Figure \ref{coadd_em}), and (4,5) two non-emitter samples binned by  UV luminosity (divided at the median luminosity $M_{\rm{UV}}=-21.4$). We measured the line EWs directly from the composite spectra; we also independently measured the same lines from the composite spectra of the four subsamples presented in \citet{shapley03}, and confirmed that our measurements return very similar values (within 0.1\AA) to their published values. These results are shown in Figure \ref{fig_ew_los} and tabulated in Table \ref{tbl_3}. In the figure, the value for our full sample is shown by a large open circle, while the two subsamples binned by Ly$\alpha$ EW are shown by filled circles. The measurements for two luminosity (non-emitters) bins are shown by filled diamonds. We also show the \citet{shapley03} and \citet{jones12} points as downward  and upward triangles, respectively. 

It is clear that Ly$\alpha$-emitting galaxies have much weaker interstellar absorption (by more than 1\AA), thus confirming the earlier studies. However, it is also true that our points for non-emitters and for the full sample lie systemically lower (i.e., stronger absorption) by $\approx 2\sigma$ than the other samples when Ly$\alpha$ EW is fixed. As a guide, we mark the $1\sigma$ range by interpolating the three Shapley subsamples (except for the highest $W_{0,{\rm{Ly}\alpha}}$ bin; Group 4). All three Jones et al. (2012) points are consistent with this relation within the errors. The fact that the measurements at $z\simeq 3$ and $z\simeq 4$ are very similar at comparable luminosities \citep[$\langle M_{\rm{UV}}\rangle =-21$:][]{shapley03,jones12} rules out that we are seeing redshift evolution of the absorbing gas. Rather, it may be that the observed discrepancy is a luminosity-dependent effect because the median luminosity of our sample is $-21.4$; i.e., 45\% more luminous). We attempted to test the luminosity-dependence by splitting the non-emitter sample into two bins according to UV luminosity (diamonds in Figure \ref{fig_ew_los}), but were unable to measure any significant difference due to the large uncertainties (each composite had only 10 galaxies). \citet{jones12} also observed a luminosity-dependent trend in their measurements, i.e., the LIS absorption  was weaker for lower-luminosity galaxies when $W_{\rm{Ly}\alpha}$ is fixed. Similarly, they reported some level of variations in LIS absorption on other galaxy parameters such as UV spectral slope, half-light radius, UV luminosity, and stellar mass (see their Figure 9). The trend that they observed is in the same direction as our observations; that more-luminous/redder/larger galaxies lie below the fiducial relation while less-luminous/bluer/smaller counterparts lie above the same relation. 

Despite relatively large uncertainties in the Jones measurements and ours, these results show a hint of changing physical conditions within the ISM that depend on galaxy properties. Since many demographic properties of galaxies correlate well with one another (luminosity, color, size, stellar mass, total mass), it is not possible to pinpoint what may be the driving factor in producing the observed trends. Nevertheless, it is still worthwhile to speculate about possible physical scenarios that may give rise to the subtle differences observed. As mentioned above, our sample contains more luminous and therefore presumably more massive galaxies \citep[e.g.,][]{lee06,lee11}. We can therefore expect these galaxies to have higher velocity dispersions on average, thus contributing to increased EWs for the interstellar absorption lines. In addition, the covering fraction of the absorbing gas may be higher in more luminous galaxies. Such a scenario would not only imply deeper (more saturated) absorption lines in more luminous galaxies, but would also explain the observed decrease in the fraction of Ly$\alpha$-emitting galaxies towards higher luminosities (as discussed in \S\ref{Lya_EW} and Figure \ref{fig_ew_muv}).  There is also circumstantial evidence that more luminous galaxies are more reddened, suggesting a higher covering fraction of dust than in less luminous galaxies. If the dust and gas distributions are similar, this also supports an increased covering fraction of gas in more luminous galaxies. Finally, it may be possible that the more luminous galaxies have ISM that are more chemically enriched. However, this last speculation is difficult to confirm without much higher quality (and higher resolution) data.

\section{Space Oddities}
%\subsection{A Large Scale Structure at $z=3.78$}\label{protocluster}
\subsection{A Protocluster Candidate at $z=3.78$}\label{protocluster}
\begin{figure*}[t]
\epsscale{1.}
\plotone{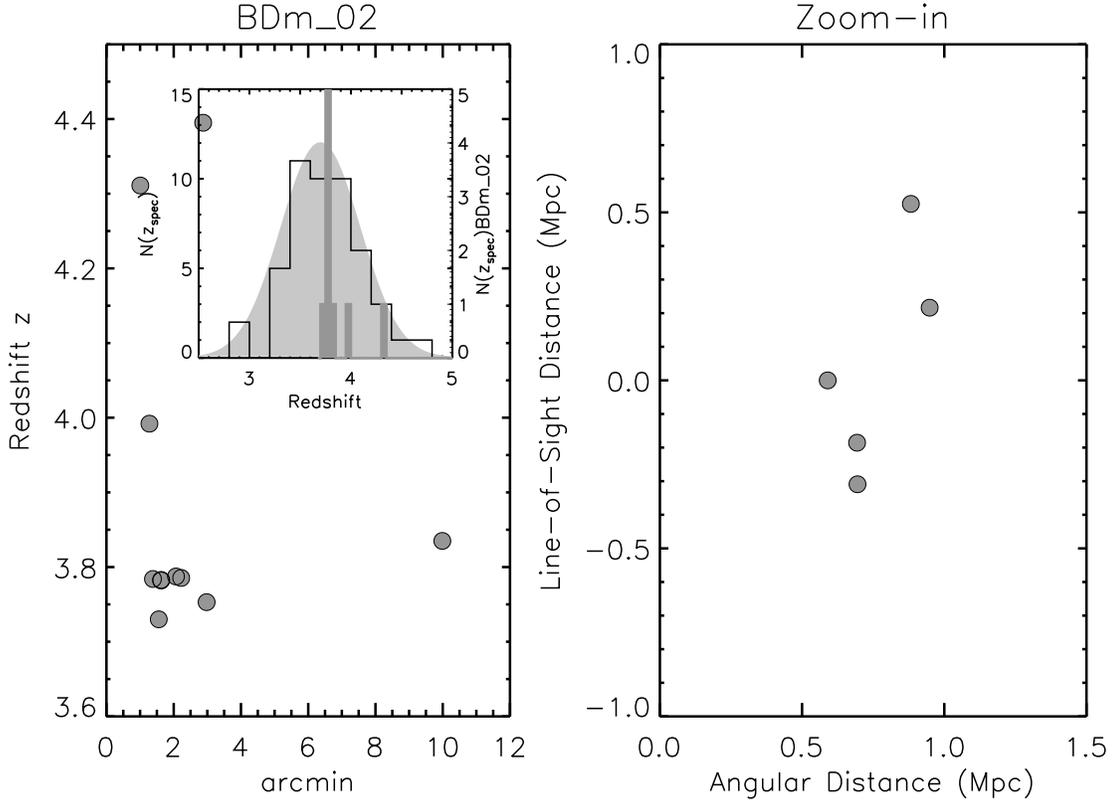}
\caption[Nz_protocluster]
{
{\it Left:} The relative positions in angular and redshift space of 11 galaxies observed on the BDm\_02 mask are shown in filled circles. The angular distance is computed from the center of the five galaxies;     RA=14$^{\rm{h}}$~31$^{\rm{m}}$~44.88$^{\rm{s}}$, DEC= 32\degr~24\arcmin~30.24\arcsec\ (J2000). 
In the inset, we show in open histogram the redshift distribution determined from the current spectroscopic sample compiled to date. The light grey shade illustrates the normal distribution with $\sigma_z=0.4$ centered at $z=3.7$. The redshift distribution of the galaxies on the BDm\_02 is shown in dark grey histogram. {\it Right:} The zoom-in on the overdensity region. The angular positions are now shown in units of Mpc (physical).
}
\label{Nz_protocluster}
\end{figure*}
The spectroscopic redshifts reveal the presence of a candidate large-scale structure at $z=3.78$. Five of the eleven galaxies on the mask BDm\_02 are identified at redshifts $3.782<z_{\rm{Ly}\alpha}<3.787$, within 1 Mpc (physical) from one another.  In Table \ref{tbl_2}, these galaxies are marked with superscript ``g''.  In Figure \ref{Nz_protocluster}, we show the relative positions of the eleven galaxies observed in the mask BDm\_02 in redshift and angular distances, and also compare with the redshift distribution determined from the current compilation of spectroscopy of our photometric candidates \citep[see][for further detail]{lee11}. As evident from the histogram, the redshift distribution on the mask BDm\_02 is highly unusual in comparison with that of the overall population in our sample. 

\begin{figure*}
\epsscale{1.}
\plotone{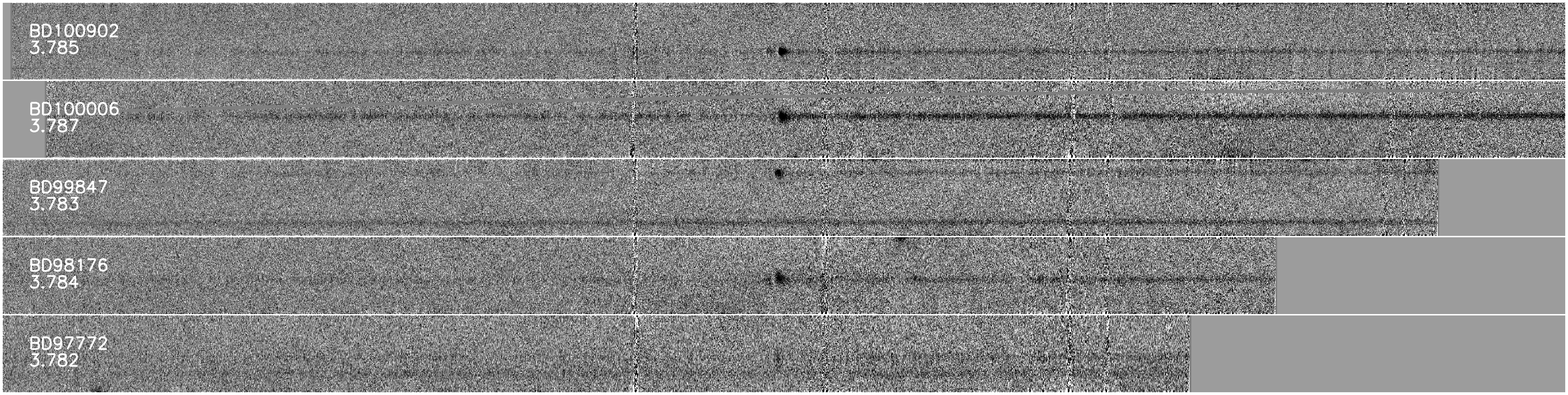}
\caption[fig_2dspec_overdensity]
{
The two-dimensional DEIMOS spectra of the five galaxies at $z=3.78-3.79$ are shown. Each spectrum is shifted to align the spectra in the observed wavelengths. The presence of Ly$\alpha$ emission is evident in all cases. The close proximity in both angular and redshift space (see Figure \ref{Nz_protocluster}) strongly suggest that these galaxies are physically associated. BD100902 and BD98176 have doubly-peaked Ly$\alpha$ emission. 
}
\label{fig_2dspec_overdensity}
\end{figure*}

We determined interstellar redshifts for four of the five galaxies, with the values ranging over $z_{\rm{IS}}=3.773-3.779$. The difference between the interstellar and Ly$\alpha$ redshifts for these sources suggest the relative velocity of $310-530$ km~s$^{-1}$. These values are comparable to the relative velocities determined for ``field'' galaxies, which we discussed in \S\ref{outflows}. Two galaxies (BD100006, BD100902) are individually detected in the SDWFS IRAC [3.6~$\mu m$] image, suggesting that they have larger stellar masses than the other three, assuming a similar star-formation history.  

All five galaxies show pronounced Ly$\alpha$ emission as can be seen in the two-dimensional spectra (Figure \ref{fig_2dspec_overdensity}). The mean and median EWs are $W_0=15.3$\AA\ and 14.2\AA, respectively, much higher than that determined for the full sample $W_0=-1.2$\AA. Similar observations were made by \citet{kuiper12} for a $z=3.13$ structure near the radio galaxy MRC 0316-257.  Three LBGs associated with the radio galaxy at $z=3.13$ have Ly$\alpha$ EWs of $W_0=14.9$, 39.3, and 17.8\AA. Based on the effective EW measured from the stacked spectrum, \citet{kuiper12} estimated $W_0=26.4\pm3.8$\AA\ for the 3 protocluster galaxies compared with $7.4\pm2.1$\AA\ for 12 field galaxies. The observed high EW is unlikely to be influenced by the spectroscopic selection effect, as all three galaxies in the MRC 0316-257 overdensity and four (out of five) of the ones in our overdensity would have been identified via absorption line features even if the Ly$\alpha$ emission was not present. Hence, these observations suggest that LBGs in the most massive potential wells may have, on average, intrinsically stronger Ly$\alpha$ emission than their counterparts in the field.  

If the  stronger Ly$\alpha$ emission observed for these galaxies is interpreted as higher SFRs, our results would imply that the galaxies in dense environments have, on average, higher SFRs than those in the field. This is the opposite of the SFR-density relation observed locally. The  ``reversal'' of the SFR-density relation  was already observed at $z\simeq 1$ \citep{elbaz07,cooper08}, where a large number of bright star-forming galaxies exist in group environments. Considering the fact that the SFR-density relation is likely produced as a result of quenching that preferentially takes place in massive systems, the reversal of such a relation is not surprising at high redshift where most galaxies are still well below the quenching ``threshold'' in masses. Accurate measurements of their stellar masses and extinction properties (UV spectral slope as a proxy for reddening in UV, or direct measurements in the far-infrared) will provide the estimates of SFR and specific SFR  thereby shedding more light on the nature of their Ly$\alpha$ emission. 

Interestingly, the only two galaxies in our sample that exhibit doubly-peaked Ly$\alpha$ emission both belong to the $z=3.78$ structure (Figure \ref{fig_2dspec_overdensity}: BD100902 and BD98176). While the frequency of similar sources at $z\simeq 2-3$ is only slightly lower $20-33$\% \citep{kulas12} than that observed in our sample, we speculate that it is possible that these galaxies may have intrinsically different physical parameters that affect the transmission of Ly$\alpha$ photons (e.g., Doppler parameter, H~{\sc i} column density, the velocity distribution of the ISM). 

Direct comparison with other known protoclusters at high redshift \citep[e.g.,][]{steidel00,ouchi05b,daddi09,capak11} is not trivial due to the varying depths and search methods of different surveys. The depth of the NDWFS data on which our selection of candidates is based allows us to probe down to $L\approx L^*$ at this redshift range, where the  galaxy surface density is significantly lower ($<0.1$ arcmin$^{-2}$) than that of typical galaxies identified as protocluster members in other studies.  For example,  the `GN20' protocluster members in the GOODS South field  are identified  down to $L\approx0.1L^*$ in their UV continuum level at the same redshift \citep{daddi09}. The surface density of $z\simeq 3.7$ photometric candidates in the GOODS South field is $\approx 9-10~ \rm{arcmin}^{-2}$ \citep[see][]{lee12a}. 

The current spectroscopy only covers an 5\arcmin~$\times$~16\arcmin\ of the area in the region,  and we are unable to determine the physical extent of the structure. Deeper imaging and more spectroscopy over a wider area is crucially needed to better quantify the uniqueness of the structure and to directly compare the physical properties and evolutionary state of its constituents with those of the field galaxies (K.-S. Lee et al., in preparation). 

\subsection{Galaxies with Multiple Interstellar Features}\label{double_abs}

\begin{figure*}[t]
\epsscale{1.}
\plotone{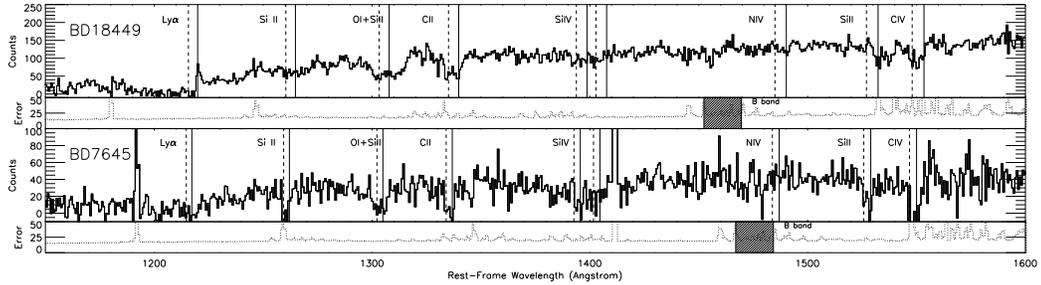}
\caption[fig_double_ISL]
{
The spectra of the  galaxies exhibiting multiple sets of interstellar absorption lines are shown.  Both error spectra are in units of counts.  The locations of prominent absorption lines are indicated by vertical  lines at two different interstellar redshifts. The interstellar redshifts derived for BD7645 (bottom) are $z=3.685, 3.679$, with the velocity offset 380 km~s$^{-1}$. On the other hand, BD18449 have $z_{\rm{IS}}=3.744, 3.726$, where the former is at rest with the Ly$\alpha$ redshift. The  velocity difference is 1140 km~s$^{-1}$, more than twice larger than the median value (\S\ref{outflows}). 
}
\label{fig_double_ISL}
\end{figure*}

Among the 41 spectroscopically confirmed LBGs at $3.2<z<4.6$, we have identified two sources, BD7645 and BD18449, that clearly exhibit two sets of interstellar absorption lines. The one-dimensional spectra of these galaxies are shown in Figure~\ref{fig_double_ISL} where prominent interstellar absorption lines are marked at two different redshifts (cyan and red  lines). 

BD7645 is continuum-dominated with Ly$\alpha$ in absorption. The absorption features are very deep with nearly zero residual intensity in Si~{\sc ii} $\lambda$1260, \c1334, Si~{\sc iv} $\lambda$1393, and C~{\sc iv} $\lambda\lambda$1548,1550 lines. The redshifts  are 3.685 and 3.679, implying the relative velocity of 380 km~s$^{-1}$, the value comparable to that reported for $z\simeq 3-4$ galaxies \citep[][see also \S\ref{outflows}]{shapley03,vanzella09}. 

BD18449, exhibits a weak Ly$\alpha$ emission in combination with deep absorption trough. We measure the net EW to be $W_{0,\rm{Ly}\alpha}=-2.2$\AA. The residual intensity is $\approx$0.5, a half of  that observed for the BD7645. The \c1334 and Si~{\sc ii} $\lambda$1526 lines are at $z=3.744$, at rest with Ly$\alpha$, while  O~{\sc i}+Si~{\sc ii} $\lambda$1303 and possibly with C~{\sc iv} $\lambda\lambda$1548,1550 lines indicate an interstellar redshift of $z_{\rm{IS}}=3.726$.  Interestingly, there is a hint of N~{\sc iv}] $\lambda$1486.5, which is sometimes seen in emission but in this case, in absorption.  There may be another component, if interpreted as a third  O~{\sc i}+Si {\sc ii} $\lambda$1303 blend, at an even higher velocity even though we cannot rule out the possibility of a chance absorber. The relative velocity of the two components in BD18449 implied by the two secure sets of interstellar redshifts is 1140 km~s$^{-1}$,  the highest  observed in our sample and twice larger than the median value $530\pm 220$ km~s$^{-1}$ (see \S\ref{outflows}).

Galaxies with multiple interstellar features are extremely rare with only a handful of reported cases; Among $\approx 3000$ galaxies at $z=1.5-3.5$ \citep[LBGs, BX/BM:][]{steideletal03,steideletal04}, there are 10 such sources identified (A. Shapley, {\it private communication}). The most well-known source is Q0000-D6 \citep[D6, hereafter;][e.g., see their Figure 9]{steidel10}, the most UV-luminous galaxy ($\mathcal{R}=22.88$) in their samples with the exception of  MS 1512-cB58, which is highly magnified. Similar to our sources, D6 shows two distinct sets of interstellar absorption lines at  $z_{\rm{IS}}=2.9561, 2.9635$, where the latter appears to be at the galaxy's systemic redshift \citep{steidel10}. D6 has a prominent Ly$\alpha$ emission which is redshifted with respect to the interstellar lines at  $z_{\rm{Ly}\alpha}=2.9692$. Another luminous source, Q1623-BX453 ($\mathcal{R}=23.38$), shares similar characteristics to D6 \citep[$z_{\rm{sys}}=2.1820$, $z_{\rm{IS}}=2.1724$, $z_{\rm{Ly}\alpha}=2.1838$:][]{law07}. 

The simplest explanation of this phenomenon is that two galaxies in close proximity are mistakenly observed as a single source. In this scenario, the continuum luminosities of the two sources should be comparable in order to produce two visible sets of absorption line features. The NDWFS $I$ and $R$-band imaging data do not reveal multiple components for the two cases here; the surface brightness profile  is consistent with a single isolated source in both cases. Thus, if the spectra are produced by  two distinct galaxies, they must lie within 0.5\arcsec\ (i.e., 3.5 kpc in physical scale) of each other, perhaps in the last stage of merger.  While we cannot rule out the possibility of witnessing two final-stage mergers, an alternate scenario would be that  the spectrum is produced by two massive rotating clumps perhaps in the process of disk formation as proposed by \citet{dekel09}. High-resolution imaging data will shed more light  on the nature of these sources.

\section{Discussion and Summary}\label{discussion}
In this paper, we have investigated the physical properties of the most UV-luminous ($L\gtrsim L^*$) star-forming galaxies at $z\simeq 3.7$ based on the rest-frame UV spectra of 41 galaxies obtained using the DEIMOS spectrograph at the W. M. Keck Observatory. The galaxies in our sample populate the exponential tail of the galaxy UV luminosity function, and have been under-represented in most previous studies due to their low surface density. They likely reside in more massive dark matter halos  and are typically found to have larger stellar masses ($\gtrsim$ a few $\times 10^{10}M_\odot$) than their less-luminous counterparts \citep{lee11}. The relatively high S/N spectra afforded by their optical brightness have allowed us to probe a variety of physical properties of individual galaxies as well as those of the population as a whole. Furthermore, we have discovered interesting classes of objects that are rarely found in lower-luminosity galaxies. These discoveries raise interesting questions about how similar the physical processes of galaxy formation are at different scales (i.e., environments, masses, luminosities). In what follows, we summarize our findings and discuss their physical implications in further detail. 

The spectroscopy presented here validate the clean and efficient selection of photometric candidates, which we have identified over a much wider area, 5.3 deg$^2$ of the Bo\"otes field, and  have used to study the general properties of LBGs \citep{lee11,lee12b}. We estimate the contamination rate to be  $\approx$11\%, mainly due to lower-redshift ($z<1$) interlopers and high-redshift QSOs (\S\ref{keck} and Figure \ref{spec}). Even in the most pessimistic case where we assume all the galaxies that could not be identified spectroscopically are not at high redshift, the contamination rate is 28\%. The median luminosity of our spectroscopic sample is 2.1 times higher than previous samples of galaxies at comparable redshifts \citep[][see Figure \ref{compare_stark}]{vanzella09, stark10}.

We find that the Ly$\alpha$ EW, i.e., the strength of Ly$\alpha$ line emission relative to the UV continuum emission, in the galaxies in our sample is generally much weaker than that found for lower-luminosity galaxies in the literature \citep{vanzella09,stark10}. The median EW of our sample is $W_0=-1.2$\AA, when we exclude galaxies that are likely not representative of our sample; these include galaxies at $z>4.2$ identified mainly via their strong Ly$\alpha$ emission despite their faint UV continuum, and five galaxies that belong to the galaxy overdensity  at $z=3.78$. We find only one galaxy with $W_0>20$\AA\ among the 41 galaxies at $3.2<z<4.2$. Furthermore, at the brightest end ($M_{1700}<-22.2$), there appears to be no galaxy with a positive Ly$\alpha$ EW. The  trend of a decreasing number of Ly$\alpha$ emitting galaxies toward higher luminosity is clear in our data, supporting a changing physical parameter that affects the escape fraction of Ly$\alpha$ photons dramatically. It is important to quantify this dependence by increasing the number of spectroscopically observed luminous ($>3L^*$) galaxies.

The most straightforward explanation of this phenomenon is that the covering fraction of neutral gas in the ISM increases with UV luminosity.  Such a picture is consistent with both the range of Ly$\alpha$ EW observed at a given luminosity, and the gradual decline in the fraction of Ly$\alpha$-emitting galaxies towards higher luminosities. Our recent {\it Herschel} observations also support this view \citep{lee12b}: by comparing the UV-derived extinction estimate (inferred by the UV continuum slope $f_\lambda\propto \lambda^{\beta}$) to a more direct measure of dust extinction (measured by the infrared-to-UV luminosity ratio), we find that 
the UV-based estimate under-predicts the true extinction by a factor of three for the most luminous galaxies (i.e., $M_{1700}<-22.2$). The two estimates can be brought into better agreement if the UV luminous galaxies follow a Small Magellanic Cloud (SMC)-like dust reddening law \citep{bouchet85} rather than the Calzetti law; the latter is often used as the standard assumption for high-redshift star-forming galaxies \citep{calzetti00}. The shallower slope for the extinction with wavelengths for the Calzetti law is generally interpreted as resulting from a clumpy ISM in the local starburst galaxies, which effectively allows more ``blue'' photons to escape without attenuation, compared to a more uniform coverage of the SMC-like ISM. Although the relative geometry of dust and neutral gas is unconstrained for high redshift galaxies, there is some evidence that they at least partially trace each other \citep{shapley03}. If the spatial distributions of dust and gas are largely decoupled, our interpretation would be invalid. We caution readers that our {\it Herschel} results are based on the photometric candidates, the majority of which are not spectroscopically verified. Even though the spectroscopy presented here seems to confirm the robustness of our sample with at most 28\% of contamination, we cannot rule out the possibility that the results may be largely influenced by lower-redshift interlopers. 

Another possibility is that the metallicity of the ISM increases with UV luminosity. The fact that interstellar absorption is generally stronger in higher-luminosity galaxies is qualitatively in agreement with this scenario. However,  the line EWs in our observations are a combined product of the column density and velocity dispersion of the absorbing gas.  Without knowledge of the intrinsic line profiles, it is difficult to disentangle these two factors and convert the measured EWs to the column density (i.e., relative abundances in the ISM).  Locally, there is evidence that the velocity dispersion is larger in more UV-luminous and/or more reddened galaxies \citep{heckman98}, which is expected if more luminous galaxies have larger dynamical masses. Measurements of elemental abundance ratios based on higher-S/N spectra may be able to constrain the relative importance of these two factors in the future. 

We investigated the relative velocity of the outflowing interstellar gas in our luminous LBG sample by measuring the difference between the redshifts of the interstellar absorption and Ly$\alpha$ emission lines, i.e., $\Delta V_{\rm{out}} \equiv V_{\rm{Ly}\alpha} - V_{\rm{abs}}$. We find that there is a correlation between the UV luminosity and $\Delta V_{\rm{out}}$ in the sense that a higher-luminosity galaxy has a higher observed ``maximum velocity'' (see \S\ref{outflows} and Figure \ref{fig_vdiff}). This implies that more UV-luminous galaxies have more powerful outflows which can potentially remove cold, neutral gas from the galaxy's potential well more effectively. Such a mechanism would have a profound effect on the subsequent formation and evolution of the galaxy. On the other hand, the gas in less powerful outflows may eventually trickle back into the galaxy, thereby only delaying the star-formation rather than preventing it entirely. 

The observed correlation in our study is qualitatively in agreement with that found by \citet{martin05} and \citet{weiner09} for local and intermediate-redshift star-forming galaxies. In contrast, observations of high-redshift LBG samples by \citet{steidel10} suggest that the outflow velocity (defined as $V_{\rm{abs}}$) is lower for galaxies with larger dynamical masses. It is possible that the large scatter in the observed velocity produced by random viewing angles has eroded any luminosity-dependent trend of the outflow velocity. The trend is subtle (e.g., Weiner et al. (2009) found $V_{\rm{out}}\propto \rm{SFR}^{0.3}$) and the wider luminosity range probed by our current study may have helped identify the trend. We also find evidence that the outflow velocity is larger for the IRAC-detected (and therefore presumably more massive) galaxies even though the low sensitivity of the current IRAC data could not provide reliable estimates of the masses for individual galaxies.

In order to study the average spectroscopic properties of the galaxies, we constructed a composite spectrum. The composite spectrum of galaxies exhibiting Ly$\alpha$ in absorption is compared to similar galaxies at $z\simeq 3$ \citep[][\S\ref{Lya_nonemitters} and Figure \ref{coadd_abs}]{shapley03}. The Ly$\alpha$ line morphology suggests that the column density and velocity distribution of the two samples are comparable. We find tentative evidence of enhanced emission in C{\sc iv} and N~{\sc v}, both indicative of stellar winds, in our spectrum compared to the $z\simeq 3$ spectrum. If confirmed, the results would have interesting implications for the IMF and metallicity of galaxies at different luminosities. We also marginally detect Ni~{\sc ii}, which is absent in the $z\simeq 3$ composite. In terms of the wind features and the relative strength of the Fe-peak element such as Ni, our spectrum may be more similar to that of MS1512-cB58 than the $z\simeq 3$ spectrum. A possible interpretation of the differences is that more luminous galaxies (as represented by the composite spectrum of our sample) may have been forming stars longer and therefore has a higher relative enrichment in the Fe-peak elements (produced by intermediate-mass, longer-lived stars) than  their less-luminous counterparts (the $z\simeq 3$ Shapley et al. composite).   

We also compared the composite spectrum of galaxies with and without Ly$\alpha$ emission at comparable UV luminosities (\S\ref{Lya_emitters} and  Figure \ref{coadd_em}). The interstellar line EWs are clearly stronger for the non-emitters, consistent with other studies in the literature. The trend likely reflects the combined effect from differing covering fractions and gas column densities of the two populations. However, we have found some evidence that there may also be intrinsic differences between the two populations.  Ni~{\sc ii} is detected in the non-emitter spectrum only, while S~{\sc v} is seen in then emitter spectrum only. The results may imply that Ly$\alpha$-emitting galaxies generally have a `younger' luminosity-weighted age than the non-emitters. The emerging picture is that the galaxy's age depends on both UV luminosity and Ly$\alpha$ EW. One simple explanation is that the covering fraction of neutral gas in the ISM increases with UV luminosity, and more luminous galaxies tend to be older. 

Two independent observations provide circumstantial evidence to this scenario. First, the non-detection of He~{\sc ii} emission in our composite, compared to its clear detection in the composite spectra of less-luminous $z\simeq 3$, $\simeq 4$ galaxies \citep{shapley03,jones12}, implies that W-R stars, which, as descendants of massive O stars, dominate the He~{\sc ii} emission at the initial period ($\lesssim 20$ Myr) of star-formation, are no longer present (\S\ref{emission_lines} and Figure \ref{compare_Si}). Second, the strengths of LIS absorption lines  are clearly higher for more UV-luminous galaxies when the Ly$\alpha$ EW is fixed \citep[\S\ref{interstellar_ew} and Figure~\ref{fig_ew_los}: see also][]{jones12}. Such effects are likely contributed by a higher covering fraction, which determines the residual intensity of the spectrum, for UV-luminous galaxies although we cannot rule out the possibility that  a higher column density and/or different kinematics of the gas are also responsible.

We have also discovered a possible large-scale structure at $z=3.78$ where five galaxies in our sample lie within 1 Mpc (physical) from one another (\S\ref{protocluster} and Figure \ref{Nz_protocluster}).  These galaxies appear to have enhanced Ly$\alpha$ emission, $\langle W_{0,\rm{Ly}\alpha}\rangle=14.2$\AA,  compared to $\langle W_{0,\rm{Ly}\alpha}\rangle=-1.2$\AA\ for the field galaxies at the same redshift range (Figure \ref{fig_2dspec_overdensity} and \ref{fig_ew_muv}). A similar trend was observed by \citet{kuiper12}, but both observations are limited by small-number statistics. We are in the process of obtaining a better census of the extent of this structure by surveying a much wider flanking region around these galaxies (K.-S. Lee, in preparation). More physical insights require more comprehensive, deeper coverage of the region, which is currently lacking. Within the hierarchical framework of galaxy formation, we argue that the discovery of a massive structure at high redshift is not at all surprising considering the unprecedented cosmic volume sampled by our candidates. Other wide-area surveys such as CFHTLS, SDSS Stripe 82, Dark Energy Survey (and, in the distant future, LSST) should be able to identify similar systems in large number, providing a clear view of how galaxy formation proceeded in the densest, most massive environments. 

Of 41 galaxies in our sample, two show evidence of multiple interstellar components that are widely separated in the velocity space (\S\ref{double_abs}, Figure \ref{fig_double_ISL}); the relative velocities between the two components are $\approx400$~km~s$^{-1}$ for one case, and $\approx1100$~km~s$^{-1}$ for the other. Galaxies with multiple interstellar components are extremely rare ($<0.3$\%) in the existing studies at lower redshift ($z\simeq 2-3$). Both galaxies appear to have a single component with a smooth surface brightness profile in the ground-based optical images, suggesting that if the interstellar features were produced by two distinct galaxies, they must be within 3~kpc from each other, perhaps in the final stages of merging. While we cannot completely rule out the possibility of a chance observation of two such events based on the small-number statistics, we also consider an alternate physical mechanism: that the interstellar features originate from giant clumps produced by  gravitational instability \citep[e.g.,][]{dekel09}. Regardless of their physical nature, the apparent high incidence rate of such systems in our sample of  $L\gtrsim L^*$ galaxies warrants further investigation to obtain better statistics.

\acknowledgments
The  data presented herein were obtained at the W.M. Keck Observatory, which is operated as a scientific partnership among the California Institute of Technology, the University of California and the National Aeronautics and Space Administration. The Observatory was made possible by the generous financial support of the W.M. Keck Foundation. We thank the staff of the W.~M.~Keck Observatory, in particular, Gregory Wirth.
The analysis pipeline used to reduce the DEIMOS data was developed at UC Berkeley with support from NSF grant AST-0071048.
KSL thanks Eros Vanzella and Tucker Jones for sharing their data and for insightful comments. The authors thank Alice Shapley for sharing her composite spectra and for useful discussions. 
The spectrum of MS1512 cB58 presented here was kindly provided by Chuck Steidel. 
This work is based in part on observations made with the {\it Spitzer Space Telescope}, which is operated by the Jet Propulsion Laboratory, California Institute of Technology. 
We are grateful to the expert assistance of the staff of Kitt Peak National Observatory where the optical and near-infrared observations of the NDWFS Bo\"otes Field were obtained. The authors thank NOAO for supporting the NOAO Deep Wide-Field Survey. The authors also  recognize and acknowledge the very significant cultural role and reverence that the summit of Mauna Kea has always had within the indigenous Hawaiian community. 

%\bibliographystyle{/Users/kyoungsoolee/publications/apj}
%\bibliography{/Users/kyoungsoolee/publications/apj-jour,/Users/kyoungsoolee/publications/myrefs}  %apj-thesis

\end{document}